\documentclass[useAMS,usenatbib]{mn2e}
\usepackage{epsfig}
\usepackage{epstopdf}

\usepackage{graphicx}
\usepackage{amssymb}
\usepackage{amsmath}
\usepackage{rotating}
\usepackage{textcomp}
\usepackage[usenames,dvipsnames]{color}
\usepackage{float,lscape}

\newcommand{\Msol}{\ensuremath{\mathrm{M}_{\odot}}}

\title[AGN-powered radio emission below 1\,mJy]{Radio-Quiet Quasars in the VIDEO Survey: Evidence for AGN-powered radio emission at $\bmath{S_{{\rm 1.4\,GHz}} < 1}$\,mJy}
\author[White et al.]
{Sarah V. White$^{1}$\thanks{sarah.white@astro.ox.ac.uk}, Matt J. Jarvis$^{1, 2}$, Boris H{\" a}u\ss ler$^{1,3}$, Natasha Maddox$^{4,5}$\\
$^{1}$Astrophysics, University of Oxford, Denys Wilkinson Building, Keble Road, Oxford, OX1 3RH, UK\\
$^{2}$Department of Physics, University of the Western Cape, Bellville 7535, South Africa\\
$^{3}$Centre for Astrophysics, Science \& Technology Research Institute, University of Hertfordshire, Hatfield, Herts, AL10 9AB, UK\\
$^{4}$Netherlands Institute for Radio Astronomy (ASTRON), PO Box 2, 7990 AA Dwingeloo, The Netherlands\\
$^{5}$Astrophysics, Cosmology and Gravity Centre (ACGC), University of Cape Town, Private Bag X3, 7701 Rondebosch, South Africa \\
}

\begin{document}

\date{Accepted 2015 January 11. Received 2015 January 11; in original form 2014 October 14}

\pagerange{\pageref{firstpage}--\pageref{lastpage}} \pubyear{2015}

\maketitle

\label{firstpage}

\begin{abstract}
Understanding the interplay between black-hole accretion and star formation, and how to disentangle the two, is crucial to our understanding of galaxy formation and evolution. To investigate, we use a combination of optical and near-infrared photometry to select a sample of 74 quasars from the VISTA Deep Extragalactic Observations (VIDEO) Survey, over 1\,deg$^{2}$. The depth of VIDEO allows us to study very low accretion rates and/or lower-mass black holes, and 26 per cent of the candidate quasar sample has been spectroscopically confirmed. We use a radio-stacking technique to sample below the nominal flux-density threshold using data from the Very Large Array at 1.4\,GHz and find, in agreement with other work, that a power-law fit to the quasar-related radio source counts is inadequate at low flux density. By comparing with a control sample of galaxies (where we match in terms of stellar mass), and by estimating the star formation rate, we suggest that this radio emission is predominantly caused by accretion activity rather than star-formation activity. 
\end{abstract}


\section{Introduction}

A galaxy is described as having an active galactic nucleus (AGN) when there is material accreting onto a supermassive black hole at the centre. Extremely luminous, unobscured AGN appear point-like at high redshifts, and so are called quasi-stellar objects (QSOs) or `quasars'. Some of these may be dust-reddened, for which there are two popular explanations: The orientation argument is that the dusty nuclear torus surrounding the accretion disc provides a large amount of extinction along certain lines of sight to the quasar (see \citealt{Antonucci1993} for a review). On the `evolution side', the quasar begins life highly-obscured, following a galaxy merger and associated starburst, but then gradually expels its dusty envelope (\citealt{Sanders1988}; \citealt{Hopkins2008}).

Such scenarios are difficult to decouple from other processes associated with either positive or negative feedback, where the star formation in the host galaxy can either be stimulated \citep{Silk2012} or truncated \citep{Croton2006} by AGN activity. Along with the peak at $z\sim 1-2$ for both accretion (e.g. \citealt{Wolf2003}; \citealt{Ueda2003}) and star-formation (e.g. \citealt{Madau1996}; \citealt{Hopkins2006}) activity, this is thought to explain, for example, why black-hole mass is correlated with the bulge mass (\citealt{Ferrarese2000}; \citealt{Gebhardt2000}; \citealt{Gultekin2009}). Much work over the past decade has revolved around understanding such feedback mechanisms, with semi-analytic models being developed to understand the observations (e.g. \citealt{Benson2003}; \citealt{Bower2006}) and help inform high-resolution simulations \citep[e.g.][]{Dubois2011,Li2014}. However, debate continues over how star-formation and accretion processes interact. 

Galaxies that have ongoing star formation, or actively-accreting black holes, will contain electrons that are accelerated in magnetic fields. This leads to radio emission being produced via synchrotron radiation (\citealt{Kellermann1988}; \citealt{Condon1992}), providing the means to measure both star formation and AGN activity at $\sim$GHz frequencies. However, for objects without obvious jets, separating star formation and accretion components is difficult at radio wavelengths. Therefore, multi-wavelength data is often used to help disentangle these two processes.

\subsection{Radio emission from AGN}

Although quasars were discovered using radio observations, only $\sim 10$ per cent have significant radio emission \citep{Hooper1996}. The remaining radio-quiet quasars (RQQs), that do not show strong jets, need to be better-studied at radio wavelengths in order to investigate their contribution to the overall radio sources counts \citep[e.g.][]{JarvisRawlings2004,Simpson2006,Padovani2014} and the underlying physical processes that may provide a different source of relativistic electrons \citep[e.g.][]{Fernandes2011,Condon2012}. 

The terms `radio-quiet' and `radio-loud' arise from the dichotomy inititally thought to exist in the AGN population (\citealt{Peacock1986}; \citealt{Xu1999}). However there are different definitions for radio loudness, historically this being a straight-forward boundary of 10$^{25}$\,W\,Hz$^{-1}$ at radio frequency 8.4\,GHz \citep{Hooper1996}. More common, and perhaps more physically meaningful, is the ratio between radio luminosity and optical luminosity \citep{Kellermann1989}. However, the distinction remains an arbitrary concept as a bimodality in radio loudness has not been convincingly proven (e.g. \citealt{Cirasuolo2003}; \citealt{Balokovic2012}). 

Whilst \citet{Ivezic2004} argue that the bimodality in the radio-to-optical luminosity ratio is genuine, others suggest it is a result of selection effects, with objects actually found across the full range of radio powers \citep{Lacy2001}. 
There is also uncertainty surrounding the exact physical mechanism that turns energy from accreted material into well-collimated jets, and why some systems show no jets at all. \citet{Lin2010} find that different radio morphologies are more dependent on the accretion rate than the galaxy's structure, with highly-extended, lobe-dominated radio galaxies having higher accretion rates than their less-extended counterparts. Furthermore, the work of \citet{Fernandes2011} indicates that there is a minimum accretion rate for a given radio power, suggesting that there is a maximum efficiency with which the black hole is able to produce jets from infalling material. Meanwhile, the influence of the spin of the black hole is still an open question. Several theoretical and observational studies suggest that it may play a role \citep[e.g.][]{BlandfordZnajek1977, Wilson1995, McLure2004, Volonteri2007, King2008, Fernandes2011}, whilst \citet{vanVelzen2013} present evidence that it could be irrelevant. 

Studies of the radio source populations as a function of radio flux density suggest that as we probe fainter radio sources, the population changes from being AGN-dominated (FRIs, FRIIs) to being dominated by star-forming galaxies \citep{Hopkins2003,Wilman2008,Condon2012}. However, there are obviously cases in which AGN are hosted in galaxies with ongoing star formation \citep[e.g.][]{Canalizo2001, Netzer2007, Silverman2009}. This is investigated further by \citet{Kimball2011} and \citet{Condon2013}, who study the radio emission from samples of optically-selected quasars, leading them to propose that the radio emission from these quasars is due to star formation within the host galaxy, rather than from the AGN. Complementary studies at other wavelengths also suggest that there is at least some ongoing star formation in quasar host galaxies. Using far-infrared observations from {\em Herschel}, \citet{Bonfield2011} find a modest correlation between accretion luminosity and star formation, and \citet{Rosario2013} show that the mean star formation rate (SFR) of quasar hosts is consistent with typical massive star-forming galaxies. 

On the other hand, \citet{Zakamska2014} use emission-line kinematics, from quasars and their host galaxies, to show that star formation in quasar hosts is insufficient to explain the observed radio emission. Instead they argue that the synchrotron emission is radiated from the shock fronts of quasar-driven outflows. Also, \citet{Simpson2006} cite radio-quiet AGN as responsible for the changing number counts, with their contribution towards the faint ($100 \leq S_{{\rm 1.4\,GHz}} < 300\,\umu$Jy) radio source population thought to be at least 20 per cent. However no distinction is made with respect to the source of the radio emission in these objects.

\subsection{Paper outline}

In this paper we extend the previous studies of \citet{Kimball2011} and \citet{Condon2013}, who use photometry and spectroscopy from the relatively shallow Sloan Digital Sky Survey (SDSS; \citealt{York2000}). We investigate the radio properties of radio-quiet quasars, selected from much deeper and narrower optical and near-infrared surveys, to determine whether there is evidence for star formation in the quasar hosts, as a function of redshift and absolute $i$-band magnitude. Section~\ref{sec:observations} describes the data used, and Section~\ref{sec:selection} explains how the sample is selected. Details of the spectroscopy, used to confirm the quasars, are given in Section~\ref{sec:spectroscopy}. We then use a template quasar spectrum to derive photometric redshifts and absolute $i$-band magnitudes, as described in Section~\ref{sec:Hewettmodel}. In Section~\ref{sec:radio} we analyse the radio emission from the quasar sample, both by stacking and by adopting a parametric description of the radio source counts for the quasars, and discuss the origin of the emission in radio-quiet quasars. Our conclusions are outlined in Section~\ref{sec:conclusions}. AB magnitudes are used throughout this paper (see Table~\ref{offsets} for conversions to Vega), and we use a $\Lambda$CDM cosmology, with $H_{0} = 70$\,km\,s$^{-1}$\,Mpc$^{-1}$, $\Omega_{m}=0.3$, $\Omega_{\Lambda}=0.7$.

\section{Data} \label{sec:observations}

Our aim is to study the radio properties of a sample of quasars, with zero contamination from inactive galaxies, as a function of luminosity and redshift. (Hereafter we describe such a sample as being `clean'.) For this we require multi-wavelength observations for the initial selection and for determining photometric redshifts. We also use follow-up spectroscopy of a subsample of our quasars, as a check on the selection to low fluxes and photometric redshift accuracy. These data are described below, and a summary of the imaging data used in this paper is provided in Table~\ref{multiwavelength}. The final area over which we select the candidate quasars is 1\,deg$^{2}$, as determined by the extent of the CFHTLS--D1 field (see Section~\ref{Sec:optical}).

\begin{table}
\centering
\caption{Multi-wavelength data used over the XMM3 tile of the VIDEO Survey, which is centred at RA(J2000) = 02:26:18, DEC(J2000) = -04:44:00. \newline CFHTLS--D1 = Canada--France--Hawaii Telescope Legacy Survey Deep field 1, covering 1\,deg$^{2}$ within the XMM3 tile. \newline VIDEO = VISTA Deep Extragalactic Observations \newline SWIRE = {\em Spitzer} Wide-area Infrared Extragalactic \newline IRAC = InfraRed Array Camera}
\begin{tabular}{@{}lccr@{}}
 \hline
   Survey name & Band & Wavelength &  Point-source sensitivity  \\
     &    &   ($\umu$m) &  (AB mag, 5$\sigma$)\\
 \hline
 CFHTLS--D1 & {\it u} & 0.38 & 27.4 \\
 CFHTLS--D1 & {\it g} & 0.48& 27.9 \\
 CFHTLS--D1 & {\it r} & 0.63 & 27.6 \\
 CFHTLS--D1 & {\it i} & 0.77 & 27.4 \\
 CFHTLS--D1 & {\it z} & 0.89 & 26.1 \\
 VIDEO & {\it Z} & 0.88 & 25.7\\
 VIDEO & {\it Y} & 1.02 & 24.5 \\
 VIDEO & {\it J} & 1.25  & 24.4 \\
 VIDEO & {\it H} & 1.65 & 24.1 \\
 VIDEO & {\it K}$_{\rm S}$ & 2.15 & 23.8 \\
 %WISE & W1 & 3.40 & 19.1 \\
 %WISE & W2 & 4.60 & 18.8 \\
 %WISE & W3 & 12.00 & 16.4  \\
 %WISE & W4 & 22.00 & 14.5 \\
 %SERVS & IRAC 1 & 3.60 & 23.1 \\
 %SERVS & IRAC 2 & 4.50 & 23.1 \\
 SWIRE & IRAC 1 & 3.60 & 22.5 \\
 SWIRE & IRAC 2 & 4.50 & 22.1 \\
 SWIRE & IRAC 3 & 5.80 & 19.7 \\
 SWIRE & IRAC 4 & 8.00 & 20.0 \\
 %SWIRE & MIPS 1 & 24.00 & 18.0 \\
 %SWIRE & MIPS 2 & 70.00 & 13.3 \\
 %SWIRE & MIPS 3 & 160.00 & 11.0\\
\hline
\label{multiwavelength}
\end{tabular}
\end{table}

\subsection{Near-infrared: VIDEO}

Our primary selection relies on near-infrared photometry and we use the 5-band near-infrared {\it ZYJHK$_{\rm S}$} data from the VISTA Deep Extragalactic Observations (VIDEO) Survey \citep{Jarvis2013}, over the XMM3 tile, which spans 1.5\,deg$^{2}$ in area. The VIDEO imaging has a typical seeing of $<$0.9\,arcsec, and a 2\,arcsec diameter aperture was used for the measurements.

\subsection{Optical: CFHTLS and VVDS} \label{Sec:optical}

The Canada--France--Hawaii Telescope Legacy Survey (CFHTLS) provides photometry in $ugriz$ over 1\,deg$^{2}$ of the VIDEO--XMM3 field \citep{Gwyn2012}. This deep field (labelled `D1') is centred at RA(J2000) = 02:25:59, DEC(J2000) = -04:29:40, and again we use a 2\,arcsec aperture for the photometry. Note that the MegaCam filters used are not identical to those of SDSS, and that the $i$ filter needed to be replaced partway through the survey. The band referred to in Table~\ref{multiwavelength} is the original $i_{1}^{\prime}$ filter, as images with the new, slightly bluer, $i_{2}^{\prime}$ filter were not used for the Deep Field stacks.

The Visible Multi-Object Spectrograph (VIMOS) VLT Deep Survey (VVDS) provides a catalogue of spectroscopic redshifts selected with $17.5 \le i \le 24.75$, with a spectral resolution of $R\approx230$ in two deep fields, VVDS--CDFS and VVDS--02h \citep{LeFevre2013}. We use data from the VVDS--02h field, which overlaps entirely with the VIDEO--XMM3 and CFHTLS--D1 fields over an area of 0.61\,deg$^{2}$.

\subsection{Radio: VLA} \label{sec:radiodata}

To investigate the level of radio emission in the quasars we use radio data, at 1.4\,GHz, from the VLA--VIRMOS Deep Field \citep{Bondi2003}. The area covered is 1\,deg$^{2}$, centred at RA(J2000) = 02:26:00, DEC(J2000) = -04:30:00, again completely contained within with VIDEO--XMM3 field. The average rms noise of the radio map is 17.5\,$\umu$Jy and the restoring beam is a $6\times 6$\,arcsec Gaussian. \citet{Bondi2003} create a catalogue of all radio components with peak flux $S_{{\rm\\ P}} > 60\,\umu$Jy ($\sim 3.5 \sigma$), resulting in 1054 radio sources, with 19 thought to be multi-component. These data have previously been cross-matched with the VIDEO survey data by \citet{McAlpine2012}.

\subsection{Mid-infrared: SWIRE}

To check our quasar selection method, we use {\em Spitzer}/IRAC imaging from the SWIRE survey that completely overlaps with the VIDEO--XMM3 field. This allows dust, associated with star-forming regions and AGN, to be traced out to $z\sim 3$ \citep{Lonsdale2003}. For the IRAC data we follow the recommendation to use `aperture 2', which gives the flux (in $\umu$Jy) as measured with an aperture radius of 1.9\,arcsec.

\section{Quasar selection} \label{sec:selection}

Many studies have used multi-colour photometry to define clean samples of quasars, free from contamination by stars and galaxies. \citet{Mortlock2012} used optical and near-infrared data, whilst \citet{Richards2006} used optical and mid-infrared data. In this paper we use a combination of optical and near-infrared for our selection, to fainter magnitudes than previous work, and check the robustness of our selection using mid-infrared data. 

In assessing whether radio emission from quasars is due to star formation in the host or from the accretion process, our key concern is to ensure a high reliability, at the expense of completeness. We therefore impose a series of cuts to maximise the reliability of our sample. \citet{Maddox2008} showed that the $J-K$ colour is very effective at selecting quasars, with the fraction being missed due to dust-reddening determined to be less than 10 per cent. This is because there is an excess in the $J-K$ colour for quasars -- even those reddened by dust -- compared to that for foreground stars. Such stars may have {\it optical} colours similar to the quasars, but near-infrared colour allows them to be distinguished \citep{Warren2000}. To exploit this, we therefore start by selecting all objects in the VIDEO $K_{\rm S}$-band brighter than the 5$\sigma$ limit of $K_{\rm S} = 23.8$, as detailed in \citet{Jarvis2013}.

\subsection{$K$-band selection} \label{sec:Kband}

\begin{figure*}
\includegraphics[scale=0.8]{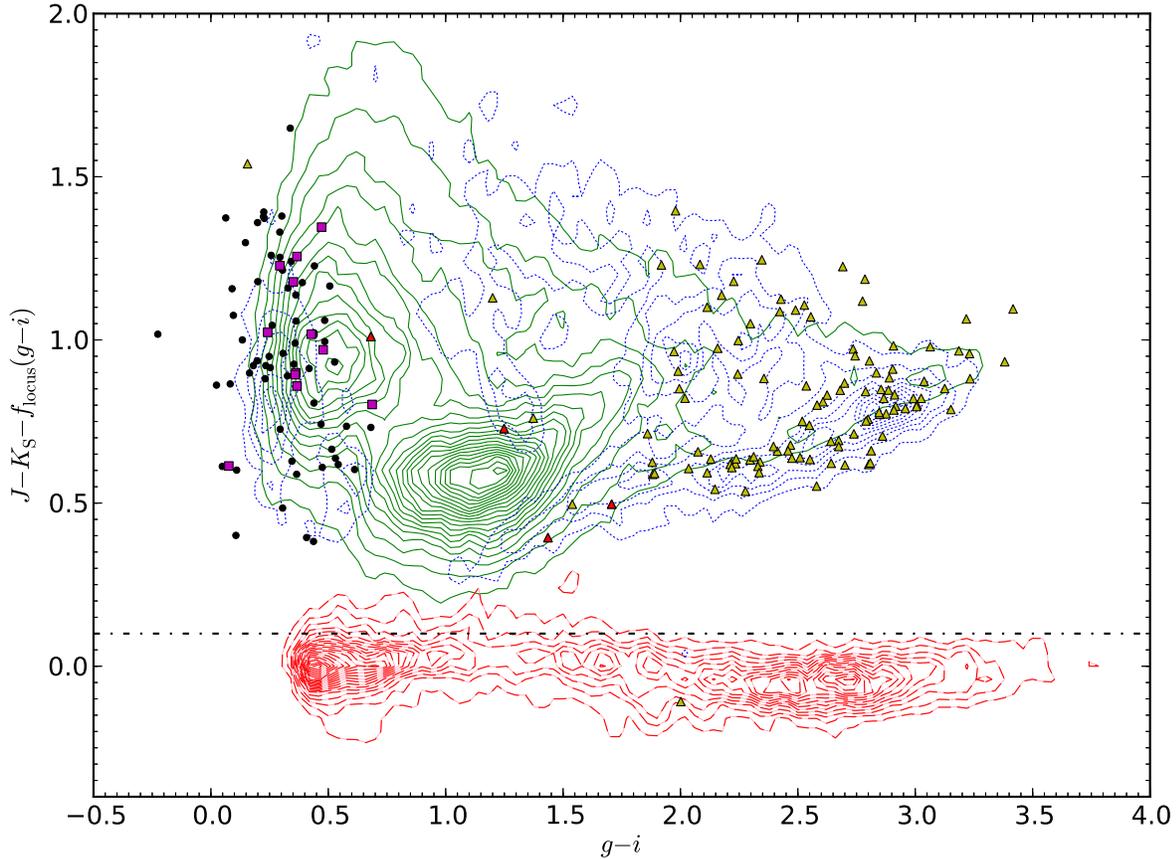}
\caption{The position of all VIDEO objects in {\it giJK$_{\rm S}$} colour space. Red contours (dashed, with interval of 2 objects) correspond to the number of objects best fit by a star template, via the {\sc Le PHARE} code (Section~\ref{sec:LePHARE}). Green contours (solid, interval = 10) are for those best fit by a galaxy template, and blue contours (dotted, interval = 2) correspond to an AGN template. The AGN and galaxy loci begin to blend into one another, whilst the objects below the black (dash-dotted) line are used to create a clean sample of stars (Fig.~\ref{pointlike}). Overplotted are objects from the `Gold' candidate quasar subsample (black dots, Section~\ref{sec:LePHARE}), including those that are cross-matched with VVDS (magenta squares, Section~\ref{sec:VVDScrossmatching}). In addition are VIDEO objects cross-matched with VVDS but lying outside the Gold selection region, either with broad emission lines (red triangles) or without (yellow triangles).  The definition of $f_{{\rm locus}}(x)$, where $x=g-i$, is as follows: for $x\leq0.3, f_{{\rm locus}}(x)=-0.6127$; for $0.3<x\leq2.3, f_{{\rm locus}}(x)=-0.79+0.615x-0.13x^{2}$; for $x>2.3, f_{{\rm locus}}(x)=-0.0632$.}
\label{BaldryVVDS}
\end{figure*}

\begin{figure}
\includegraphics[scale=0.42]{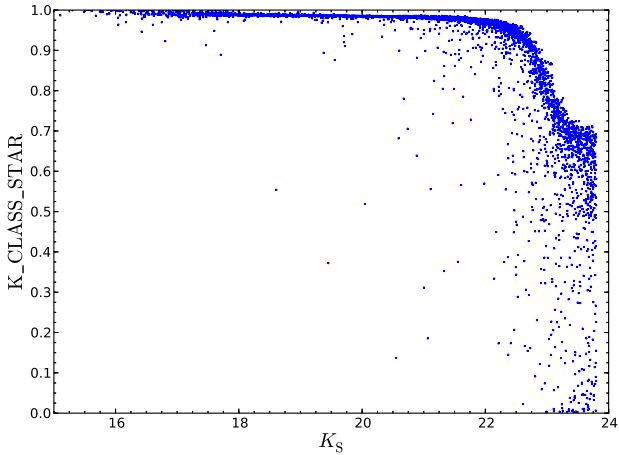}
\caption{The variation in K\_CLASS\_STAR, an indicator of compactness (1 = point-like), for a clean sample of stars. Each of these objects are $K_{{\rm S}}$-band selected ($K_{\rm S}\le$ 23.8) in the VIDEO catalogue, and have stellar colours using the locus shown in Fig.~\ref{BaldryVVDS}}.
\label{pointlike}
\end{figure}

We expect the vast majority of quasars to outshine their host galaxies in the $K_{\rm S}$-band, and therefore impose a morphology cut based on the {\sc SExtractor} K\_CLASS\_STAR parameter \citep{Bertin1996}. This is a measure of the compactness of a source on a scale of 0 to 1, with 1 representing a point-like appearance. To check the effectiveness of K\_CLASS\_STAR as a morphology indicator, we select stars from the VIDEO catalogue using {\it giJK$_{\rm S}$} colours (Fig.~\ref{BaldryVVDS}). Following the method of \cite{Baldry2010}, \cite{Jarvis2013} show that the region of colour space defined as $J-K_{\rm S}-f_{{\rm locus}}(g-i) < 0.1$ provides a clean stellar sample, free from contamination by galaxies. Fig.~\ref{pointlike} shows the value of K\_CLASS\_STAR as a function of $K_{\rm S}$-band magnitude for these objects, and we see that our point-source classification is robust down to $K_{\rm S}=22.8$. Therefore imposing a restriction on the morphological parameter, K\_CLASS\_STAR $>$ 0.9, effectively means that we should eliminate objects with $K_{\rm S} \geq 22.8$. 

As a further check, since we will rely on this morphology parameter for quasars rather than stars, we create a set of simulated AGN. This is done by adding a point source, representing the output from the quasar nucleus, to the image of a model host galaxy. The latter are simulated using the methodology of \citet{Haeussler2007}. For numerous steps in the total $K_{\rm S}$-band magnitude, different values of AGN magnitude ($m_{\rm AGN}$), galaxy magnitude ($m_{\rm galaxy}$), and effective radius for the host ($r_{\rm eff}$) are assigned (see Appendix~\ref{sec:completeness}). {\sc SExtractor} is then used on the simulated images, outputting the K\_CLASS\_STAR value for each object. Next, for each subset of the resulting catalogue -- defined by a particular combination of total magnitude ($K_{\rm S}$), $m_{\rm AGN}$ and $m_{\rm galaxy}$ -- the fraction of objects with K\_CLASS\_STAR $>$ 0.9 is calculated (Table~\ref{completeness}). This provides an estimate of our detection completeness, with respect to the AGN.

A simulated AGN that has been correctly classified as point-like will have a K\_CLASS\_STAR value close to 1. Our results show that for the brightest objects, with $K_{\rm S} < 19.0$, the AGN must account for at least 90 per cent of the total flux if over 90 per cent are to have K\_CLASS\_STAR $>$ 0.9. This corresponds to the AGN needing to outshine its host galaxy by over two orders of magnitude. The reason for this is that the host galaxies belonging to such bright systems are themselves bright, and so their extended appearance influences the K\_CLASS\_STAR measurement. However, we do not expect this to greatly bias our sample as very few quasars in our sample have $K_{\rm S} < 19.0$. Over the more populated range of $19.0 \leq K_{\rm S} \le 21.5$ (due to the shape of the quasar luminosity function and depth of our data), we are able to recover over 90 per cent of the simulated AGN provided that their flux is $\geq 50$ per cent of the total flux. We find that the distribution of $r_{\rm eff}$ for the host galaxy over this regime has no influence on the K\_CLASS\_STAR values.% (i.e. $m_{\rm AGN} \leq m_{\rm galaxy}$) %For interest, 4 out of the final 74 qsos have K<19

To achieve $\geq 90$ per cent completeness over $21.5 \leq K_{\rm S} \le 22.5$, we find that the AGN's contribution to the total flux must rise again to at least 70 per cent. This means that we should be aware of possible contamination from the host galaxy, as a consequence of using the K\_CLASS\_STAR $>$ 0.9 cut across this magnitude range. For total magnitude $K_{\rm S} > 22.4$, no level of contribution from AGN to the total flux ($f_{\rm AGN}/f_{\rm total}$) allows us to attain even 50 per cent completeness. Therefore we impose a cut of $K_{\rm S} \leq 22.4$, in addition to using K\_CLASS\_STAR $>$ 0.9 as a selection criterion, which ensures that the morphology of the candidates is reliable. Whilst this means that we may be biased towards the brightest quasars and faintest host galaxies, such a conservative approach is sufficient for the purpose of this paper. Indeed, it strengthens the conclusions we draw regarding the radio emission, as explained in Section~\ref{sec:discussion}. 

\subsection{Photometric fitting} \label{sec:LePHARE}

\begin{figure*}

\includegraphics[scale=0.8]{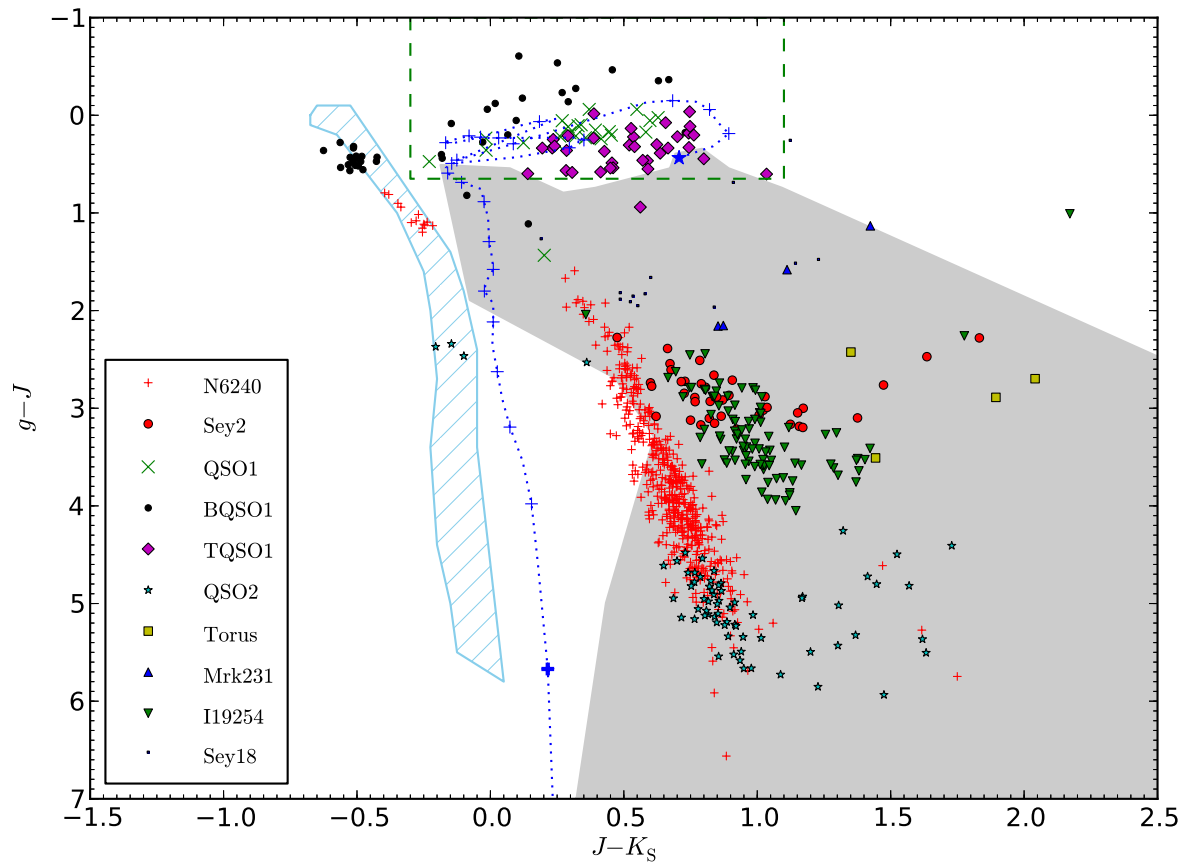}
\caption{The positions of point-like objects (i.e. those with K\_CLASS\_STAR $>$ 0.9) in {\it gJK$_{\rm s}$} colour space, following \citet{Maddox2008}. Each of these data points has photometry that is best fit by one of 10 AGN templates (see legend, and their description in Section~\ref{sec:LePHARE}). The light-blue, hatched region represents the stellar locus, occupied by VIDEO objects for which a stellar template provides the best fit to the photometry. The blue dotted track is a model for how a quasar with no host galaxy contamination evolves with redshift, lying between the stellar and galaxy loci (to the left and right, respectively). The redshift range of this track is $0.0 \leq z < 5.0$, with a blue star marking $z=0.0$ and blue `+' markers every $\Delta z = 0.2$ (with a bold marker denoting $z=4.8$). The grey shaded region is the area covered by evolution tracks for the photometric fitting code's library of galaxy templates. The redshift range of these galaxy tracks is $0 \leq z < 6$, with colours evolved for each step of $\Delta z=0.04$ taken. The region used to select `Gold' candidate quasars is demarcated by a green dashed line.}
\label{all}
\end{figure*}

As stated previously, optical bands alone have in the past been used to define a colour space for quasar selection (e.g. \citealt{Richards2001}). However their success in separating the regions occupied by stars and quasars diminishes at high redshifts, and they are especially affected by dust-reddening. Instead we follow the method of \citet{Maddox2008} and use {\it gJK$_{\rm S}$} colour space (Fig.~\ref{all}). 
We also use full spectral energy distribution (SED) modelling, as inactive galaxies may still be present amongst our potential quasar candidates, contaminating our sample. Fitting SED templates to the VIDEO--CFHTLS--D1 data was carried out using {\sc Le PHARE} \citep{Ilbert2006}.

There are 62 galaxy templates, based on the SEDs compiled by \cite{Coleman1980} and those of starburst galaxies. In combination with different extinction law values, these produce 187 sets of model photometry that are then redshifted. A range of extinction values are also used in conjunction with the stellar library, consisting of 254 templates, and the 10 empirical SEDs that make up the AGN library. 

These AGN templates are taken from the SWIRE template library \citep{Polletta2007}. Two of them, `Sey18' and `Sey2', represent moderately-luminous Seyfert galaxies. The `I19254' template is from IRAS19254-7245, a starburst/ULIRG with Seyfert-2 galaxy properties. Another ULIRG is included in the form of `Mrk231', which has a heavily-obscured quasar at the centre. Like `I19254' and the `N6240' AGN template, its SED is an AGN-starburst composite. 

\citet{Hatziminaoglou2005} used an optically-selected sample of 35 SDSS/SWIRE Type-1 quasars to construct the three Type-1 quasar templates (`QSO1', `BQSO1', `TQSO1'), via a quasar composite spectrum in addition to rest-frame infrared data. The difference between them is their IR/optical flux ratios, with `BQSO1' representing the bottom quartile and `TQSO1' the top quartile. Finally, the `QSO2' and `Torus' templates are typical SEDs for Type-2 quasars, with the second of these having greater dust obscuration. We determine the best fit for both redshift and extinction using each of the SED libraries (`star', `galaxy' and `AGN'). 

Fig.~\ref{all} shows the $gJK_{\rm S}$ colour space for the point sources in the VIDEO $K_{\rm S}$-band selected catalogue where one of the ten AGN SEDs provides a better fit to the photometry than the galaxy or star templates. Point-like objects that are best-fit by a stellar template are predominantly found in the light-blue, hatched region. Therefore the data points that lie within this region represent objects with stellar $gJK_{\rm S}$ colours, despite their photometry being better-modelled by an AGN SED. The blue, dotted line in Fig.~\ref{all} is the evolution track for a model quasar that has no host galaxy contamination, and follows a path that lies between the star and galaxy regions.

We also show the broad region occupied by the galaxy tracks that are used by {\sc Le PHARE} (shaded grey in Fig.~\ref{all}). Some of the AGN template-fitted data points are well-separated from the galaxy region, whilst others have greater overlap. Those that remain `clear' from the galaxy locus, within the region demarcated by a green dashed line (defined by the equations listed in Appendix~\ref{sec:equations}), are subsequently referred to as the `Gold candidate quasar sample'. We use the descriptor `Gold' simply to distinguish the candidates that we are confident of being quasars. (Additional samples, labelled `Silver' for example, could be constructed using more-relaxed selection criteria.) The Gold sample comprises 75 quasars, with one later being removed as the result of an absolute $i$-band magnitude cut (Section~\ref{sec:luminosities}). We deem the slight overlap with the galaxy tracks (shaded region) as acceptable given the photometric uncertainties. The selected objects are best-fitted by Type-1 quasar models: `QSO1', `BQSO1' and `TQSO1', and we later see where they lie in mid-infrared colour space (Section~\ref{sec:MIR}). Valuing reliability over completeness, we chose not to extend the selection region downwards to include a few extra points fitted by Type-1 quasar models. This is because, with their positions lying well within the shaded region, there is a greater chance of their $gJK_{\rm S}$ colours being confused for galaxies. Similarly, the selection region is not extended bluewards to include the BQSO1 objects around $J-K_{\mathrm{S}}=0.5$, $g-J=0.5$, as doing so may introduce contamination by stars.

A similar method is used by \citet{Maddox2012} to construct a sample of quasars, whose {\it K}-band magnitude range is 15.8 $\le K \le$ 18.4. They are able to achieve $> 95$ per cent efficiency and $> 95$ per cent completeness with respect to known SDSS quasars. In addition, both their and our methods incorporate information from SED fitting. Doing so enables us to be conservative in our selection, resulting in a clean sample of candidates. However, in contrast, our candidates are much fainter due to the depth of the photometry used, spanning 18.4 $\le K_{\rm S} \le$ 22.4. The light from the host galaxy could lead to greater contamination as we go deeper, but our selection (via morphology and colour information) ensures that the quasar is dominating the total flux from the system.

As a result of our selection criteria, we actually have very good completeness with respect to {\it unobscured} quasars, as the majority of Type-1 sources lie in the selection region (demarcated by the green, dashed line in Fig~\ref{all}). The SEDs that we have excluded are for composite systems or obscured AGN. This means that our sample is not representative of the entire quasar population, and so we emphasise that our investigation accounts for the radio emission from Type-1 quasars only. Also, the use of near-infrared data may lead to a bias towards bright quasars residing in faint host galaxies. This is because a host that is bright in the $K_{S}$-band may have an extended appearance in the imaging data, causing the object to be eliminated by our morphological cut. This potential bias is discussed in the context of our final results in Section~\ref{sec:discussion}.

\subsection{Mid-infrared colour}\label{sec:MIR}

As an additional check on the validity of our selection, we use {\em Spitzer} data to investigate where the Gold candidate quasars lie in mid-infrared colour space. Dust extinction is not a problem when measuring emission in the mid-infrared, and so both unobscured and obscured AGN can be detected. However, we note that our Gold candidates are biased towards the {\it unobscured} population, given the photometric fit to Type-1 quasar SEDs. To check what proportion are indeed active-galaxy candidates, we use the work of \citet{Lacy2004} and \citet{Stern2005}.

\citet{Lacy2004} use the 8.0\,$\umu$m/4.5\,$\umu$m ratio versus the 5.8\,$\umu$m/3.6\,$\umu$m ratio to separate stars, low-redshift galaxies, SDSS quasars and radio-selected quasars. Within this colour space they find that there are two distinct sequences, the quasars belonging to one and the remaining objects to the other. We refer to the region that contains the quasar sequence as the `Lacy wedge'. 

The tendency of AGN to be redder than galaxies in the mid-infrared is also exploited by \citet{Stern2005}, who use a $[3.6]-[4.5]$ versus $[5.8]-[8.0]$ colour-colour space. The combination of restricted $[5.8]-[8.0]$ colours observed for powerful AGN (due to a lack of strong PAH emission) with their $[3.6]-[4.5]$ colour (much redder than that of low-redshift galaxies) leads to another clear selection region. We call this the `Stern wedge'. 

\begin{figure*}
\includegraphics[scale=0.55]{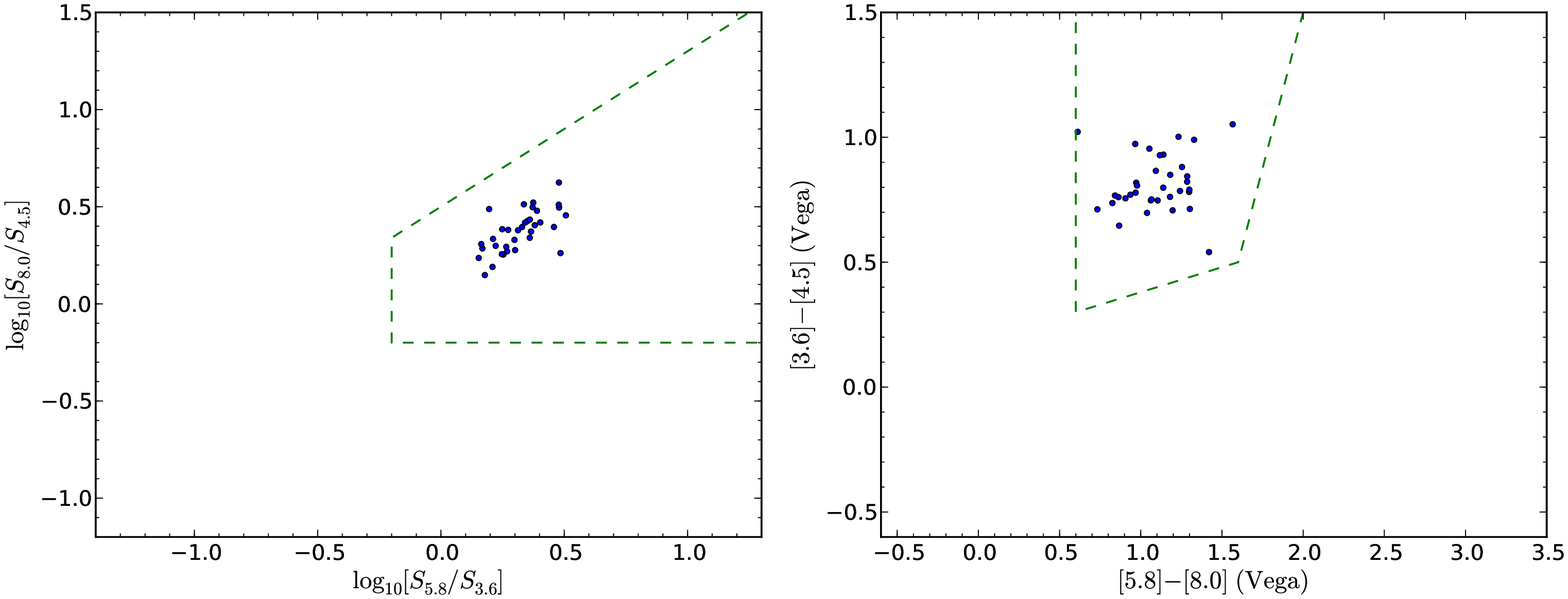}
\caption{Gold objects that have mid-infrared data are shown, following \citet{Lacy2004} (left) and \citet{Stern2005} (right). The dashed lines outline the regions that these authors use to select active-galaxy candidates.}
\label{LacyandStern}
\end{figure*}

We find that 62 of our 75 Gold quasar sample have cross-matches in the SWIRE catalogue. Due to the depth of the mid-infrared data, only 35 of these have detections in each of the 4 bands, IRAC 1-4. Hence we use this as a consistency check on the brighter candidates, rather than an additional step in the selection. Fig.~\ref{LacyandStern} shows that all of the Gold sample (with detections in all 4 bands) lie within the Lacy and Stern wedges. This provides further evidence that our quasar selection method is robust.

\section{Spectroscopy} \label{sec:spectroscopy}

In this section we use spectroscopic data to carry out a further check of the robustness of our sample selection, in addition to obtaining accurate redshift information. Existing spectroscopy comes from the Visible Multi-Object Spectrograph (VIMOS) VLT Deep Survey (VVDS), over the XMM3 tile within the VIDEO field. This is complemented by new spectroscopy for objects belonging to our Gold sample, using the Southern African Large Telescope (SALT). We discuss how the spectroscopic redshifts compare to the photometric redshifts obtained from the template fitting in Section~\ref{sec:Hewettz}.

\subsection{VVDS data} \label{sec:VVDScrossmatching}

We cross-matched all of the objects that had best-fit SEDs of an AGN (obscured, unobscured, or Seyfert-like) with the VVDS, using a matching radius of 1\,arcsec. This identifies 127 objects with spectroscopic redshifts over the 0.61\,deg$^{2}$ covered by the VVDS--02h field, twelve of which are in our Gold sample. Each of these 12 objects have broad emission lines, again showing that we have a good basis for selecting quasars. They also lie in the quasar region of $giJK_{\rm S}$ colour space (magenta squares in Fig.~\ref{BaldryVVDS}), as identified by \citet{Baldry2010}.

The remaining 115 cross-matches lie in the galaxy locus of both $giJK_{\rm S}$ space (triangles in Fig.~\ref{BaldryVVDS}) and $gJK_{\rm S}$ space (Fig.~\ref{all}). According to the VVDS redshift flags, 7 of them have broad emission lines, meaning that they are more likely to be quasars. However, inspection by eye reveals that this is actually the case for only 5 of the objects, one of which is a Broad Absorption Line (BAL) quasar. We also confirm that reliable redshifts have been obtained for each of these 5. They appear as red triangles in Fig.~\ref{BaldryVVDS}, with one lying beyond the plot range of the figure. Their colours are redder than the Gold quasar candidates, and lie in the region of $gJK_{\rm S}$ colour space covered by galaxy evolution tracks, hence their exclusion from our sample. These extra quasars highlight the incompleteness of our Gold sample, although we emphasise our aim for a highly-reliable selection method instead. The 110 cross-matched objects that show no broad emission lines, yet are best-fit by an AGN template, may be obscured AGN.

\subsection{SALT spectra}\label{sec:SALTspectra}

13 quasars from our Gold sample were observed between 4th November 2012 and 6th January 2013, using the Robert Stobie Spectrograph \citep{Kobulnicky2003} on SALT \citep{Buckley2006}. We used the PG0900 grating with the PC03850 filter, to obtain coverage over 4500--7500\,\AA. A 2 arcsec slit provided adequate resolution (10\,\AA) for measuring redshifts from broad emission lines. Three exposures were made for each target to enable cosmic ray rejection.

The standard procedures for reducing SALT data were performed with the {\sc IRAF} software. For flux calibration, LTT 1020 was used as the standard star. The resulting fully-reduced spectra, extracted from the calibrated images, are presented in Appendix~\ref{sec:allspectra}. They correspond to the 8 Gold candidates for which it was possible to determine reliable spectroscopic redshifts. These are discussed further in Section~\ref{sec:Hewettz}. Strong emission lines are present in all of the spectra, with C{\sc iii}]$_{1909}$ being the most common. Also present is either a C{\sc iv}$_{1549}$ or Mg{\sc ii}$_{2799}$ broad line. Spectra for the other 5 objects observed by SALT were poor, due to being badly affected by weather and technical difficulties.

\section{The quasar sample} \label{sec:Hewettmodel}

\subsection{Photometric redshifts} \label{sec:Hewettz}

The photometric redshifts, fitted using {\sc Le PHARE} (Section~\ref{sec:LePHARE}), were compared to the spectroscopic redshifts of the Gold objects, obtained with the VVDS and SALT. The majority are in good agreement (see Table~\ref{tab:photz}), but there are a few cases where the photometric fitting severely underestimated the true redshift. Using the normalised median absolute deviation in $\Delta z/(1+z)$, where $\Delta z = (z_s - z_p)$,  we find $\sigma_{\Delta z/(1+z)} = 0.082 $ and four catastrophic outliers, where we define a catastrophic outlier as having $|\Delta z|/(1+z) > 0.15$ \citep{Ilbert2006}. Such inaccuracies prompted us to seek an alternative method for determining photometric redshifts.

We therefore used a different quasar template that has no host galaxy contribution (Hewett et al., in prep.). The colours from this template were evolved over the redshift range $0<z<7$ in steps of 0.1, and we also included 0.2\,mag of uncertainty to the quasar template colours to account for the variation in emission-line strength and the intrinsic quasar spectral shape. This was added in quadrature to the measured photometric uncertainties. The best fit was found through $\chi^{2}$ minimisation, and we name the corresponding redshift `Colour-$z$'. Colour-$z$ values calculated for the objects with spectroscopic redshifts are presented in Table~\ref{redshifts}.

\begin{table}
\centering
\caption{Summary of redshifts for the QSOs spectroscopically confirmed by the VVDS, and those observed by SALT (except 5 for which a redshift could not be determined). * denotes a VVDS object assigned an incorrect redshift; the redshift given is our measurement (see text for details).}\label{tab:photz}
\begin{tabular}{@{}llccr@{}}
 \hline
 R.A. & Dec. &{\sc Le PHARE}  & Colour & VVDS/ \\
 (hms)&(dms)&$z$& $z$& SALT $z$\\
 \hline
02:25:25.68	&-04:35:09.6&	0.4&	2.1&	2.1 \\   	%1060199 	(VVDS)	
02:25:45.53	&-04:34:45.6&	1.7&	1.7&	1.7*\\	%1061136	(VVDS)	
02:25:50.38	&-04:33:24.6&	2.8&	2.5	&2.7\\	%1064151	(VVDS)
02:25:52.16 &    -04:05:16.1& 1.5 & 1.4 & 1.4 \\ %ID=1132702 (SALT)
02:26:09.62	&-04:24:38.0&	2.6&	2.8	&2.7\\	%1086539	(VVDS)
02:26:12.64 &    -04:34:01.4& 2.2 & 0.7 & 2.3 \\ %ID=1062840 (SALT), catastrophic outlier
02:26:18.59 &    -04:11:01.1& 0.4 & 1.9 & 2.0 \\ %ID=1118527 (SALT)
02:26:22.15	&-04:22:21.8&	2.0&	1.8&	2.0\\	%1092183	(VVDS)
02:26:24.64	&-04:20:02.4&	2.0&	0.7	&2.2\\	%1097378	(VVDS), catastrophic outlier
02:26:33.31 &    -04:29:47.8& 2.4 & 2.4 & 2.1 \\ %ID=1073715 (SALT)	
02:26:39.83&-04:20:04.4& 1.9 & 1.5 & 1.6 \\ %ID=1097196 (SALT)
02:26:52.14 &    -04:05:57.1 & 1.4 & 1.4 & 1.4 \\ %ID= 1131001 (SALT)
02:27:07.54	&-04:32:03.2&	1.6&	1.5	&1.5*\\	%1067992	(VVDS)
02:27:09.03 &    -04:55:10.1& 3.0 & 2.7 & 2.7 \\ %ID=1011226 (SALT)
02:27:33.99	&-04:25:23.5& 1.8&	1.5&	1.6\\	%1084813	(VVDS)	
02:27:36.93	&-04:26:31.5&	0.6&	1.8	&1.8\\	%1081826	(VVDS)	
02:27:38.99	&-04:09:40.8&	1.4&	1.3	&1.4\\	%1122090	(VVDS)
02:27:40.55 &    -04:02:51.1 & 2.4 & 2.6 & 2.6 \\ %ID=1138526 (SALT)	
02:27:47.34	&-04:27:53.7&	0.2&	2.4	&1.1\\	%1078470	(VVDS), catastrophic outlier
\hline
\label{redshifts}
\end{tabular}
\end{table}

\begin{figure}
\includegraphics[scale=0.42]{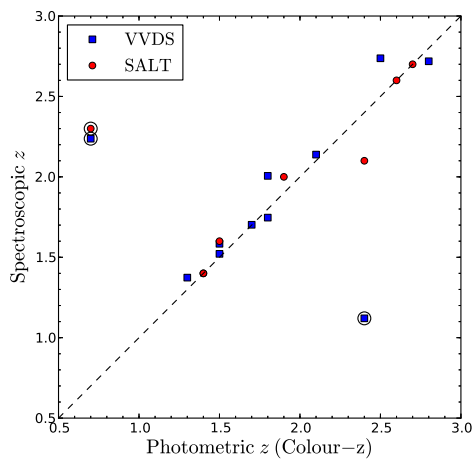}
\caption{Spectroscopic redshifts, from SALT and VVDS, showing the accuracy of Colour-$z$ as a photometric redshift, including the catastrophic outliers (circled, see text for details).} %$\sigma_{\Delta z/(1+z)} = 0.046 $.}%Two objects are at (1.4,1.4)
\label{VVDSSALTz}
\end{figure}

Fig.~\ref{VVDSSALTz} shows the correlation between the spectroscopic redshift, from both SALT and VVDS, and our photometric redshift. Two of the VVDS redshifts were highly discrepant, quoted to be 0.1 and 0.2 (Table~\ref{redshifts}). Inspection of their spectra showed that a Mg{\sc ii}$_{2799}$ line had been mis-identified as a H$\alpha_{6563}$ line. We therefore corrected the redshifts and used these values for Fig.~\ref{VVDSSALTz}, which placed them in line with their corresponding Colour-$z$ value. Using all of the objects with spectroscopic redshifts, the accuracy of Colour-$z$ is $\sigma_{\Delta z/(1+z)} = 0.046$. Out of the 19 objects we find there to be 3 catastrophic outliers (circled in Fig.~\ref{VVDSSALTz}). We investigated whether confusion between the Mg{\sc ii}$_{2799}$ line and the C{\sc iv}$_{1549}$ line led to the catastrophic outliers with Colour-$z=0.7$, and similarly whether a C{\sc iii}]$_{1909}$ line was being confused for a Mg{\sc ii}$_{2799}$ line, resulting in the third outlier (with Colour-$z=2.4$). However, in each case, the absence/presence of other expected emission lines confirmed the previously-determined spectroscopic redshifts.

Therefore our Colour-$z$ photometric redshift estimates perform well for the majority of quasars observed by the VVDS and SALT. However, a single spectrum is not enough to describe the variety found in the Type-1 quasar population, and for this reason we still find a number of outliers. As such, we cannot simply assume these best-fit values for the remainder of the photometric sample, for which we do not have spectroscopy. Instead we note that each of these outliers have a double-peaked probability distribution function (PDF) produced by the photometric fitting, with the secondary peak overlapping the spectroscopic redshift. This prompted us to generate 1000 Monte Carlo redshifts per Gold quasar, according to the full PDFs calculated for the Colour-$z$ redshifts. These redshifts are used for the distribution in absolute $i$-band magnitudes (Section~\ref{sec:luminosities}) and the subsequent analysis of the radio emission associated with the quasars. This therefore allows our final results to take the redshift uncertainties into account.

\subsection{Absolute {\it i}-band magnitudes} \label{sec:luminosities}

\begin{figure}
\includegraphics[scale=0.4]{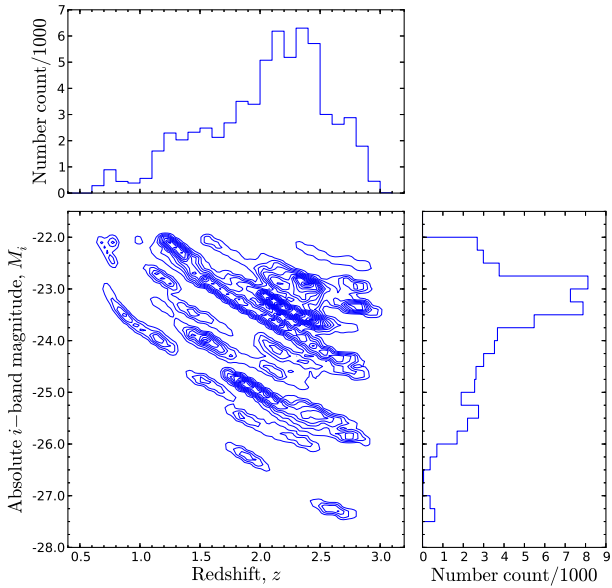}
\caption{The distribution of photometric redshifts (Colour-$z$) and {\it K}-corrected absolute {\it i}-band magnitudes for the final 74 Gold quasars with $M_{i}\leq-22$ (contour interval=10), determined using the full redshift probability distributions.}
\label{RedshiftsAbsimag}
\end{figure}

The same quasar spectrum (Hewett et al., in prep.), in addition to the Colour-$z$ values determined in the previous subsection, is used to calculate the $K$-corrected absolute $i$-band magnitudes ($M_{i}$) of our quasar sample. We also use the redshift probability distributions to calculate the probability distribution for $M_{i}$ per quasar. We then impose the criterion $M_{i}\leq-22$, a magnitude cut that approximately splits quasars from Seyfert galaxies \citep[e.g.][]{Schneider2003}. The objects that survive this cut are presented in Fig.~\ref{RedshiftsAbsimag} and used for the remainder of the paper. We find one quasar has {\it no} part of its redshift probability distribution with $M_{i}<-22$, and so this object is removed from the sample. The properties of the final 74 quasars are given in Table~\ref{74quasars}. 

\begin{figure}
\includegraphics[scale=0.42]{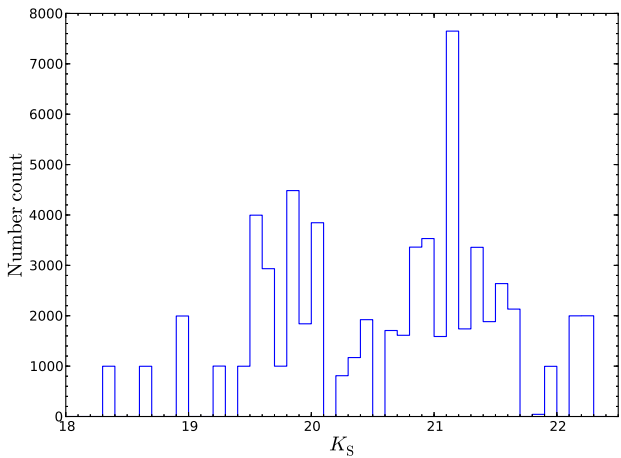}
\caption{The $K_{{\rm S}}$-band magnitude distribution for the final 74 Gold quasars, for those that have $M_{i}\leq-22$ generated from sampling the redshift probability distribution.}
\label{final74quasarsKmag}
\end{figure}

61 of these 74 objects are beyond $z=1$ and brighter than $M_{i}=-22$, demonstrating the effectiveness of VIDEO+CFHTLS in detecting very faint quasars. However, some of the objects may have an incorrect $M_{i}$, arising from an incorrect Colour-$z$. This is why it is important to note that we take this into account via our sampling of the photometric-redshift probability distribution. For example, the fraction of simulated objects that have $M_{i}\leq-22$ is used to weight the Gold quasars when analysing their radio emission in the next section. In Fig.~\ref{final74quasarsKmag} we also show the $K_{\rm S}$-band magnitude distribution of the final 74 quasars. Note that the majority are indeed in the range $19.0 \leq K_{\rm S} \le 21.5$, as previously mentioned in Section~\ref{sec:Kband}.

To understand the proportion of the total quasar population we have selected with our criteria, we use the quasar luminosity functions modelled by \citet{Croom2009}\footnote{We note that the value given in table 4 of \citet{Croom2009} should be $\log_{10}(\Phi_{0}^{*})=-5.99\pm0.07$ (S. Croom, private communication).}. These are integrated over the redshift range of the final Gold sample, $0.5\leq z\leq 3.1$ (Fig.~\ref{RedshiftsAbsimag}), for objects brighter than $M_{g}(z=2)=-22.8$, which corresponds to our cut of $M_{i}(z=0)=-22.0$. Using the luminosity dependent density evolution (LDDE) and luminosity evolution + density evolution (LEDE) models as lower and upper limits, respectively, these luminosity functions predict there to be between 338 and 381 quasars over 1\,deg$^{2}$. We are therefore approximately 20 per cent complete. 

\section{Radio emission from radio-quiet quasars}\label{sec:radio}

\subsection{Radio flux-density measurement} \label{sec:extraction}

To obtain the radio properties of the RQQs in our sample, we initially performed a stacking analysis. We used 1.4\,GHz flux densities extracted from the VLA--VIRMOS Deep Field map \citep{Bondi2003} on a single-pixel basis, at the positions of the quasars. Due to the conversion from right ascension and declination to an integer pixel number, the positional error of the quasar associated with the chosen pixel is 1.5\,arcsec for both co-ordinates. This error is comparable to the angular size of the objects in question, but deemed negligible given the 6\,arcsec resolution of the data. 

The mean and median radio flux-density values obtained at the positions of the Gold objects are shown in Table~\ref{fluxes}, noting that the radio map has an average rms noise of 17.5\,$\umu$Jy \citep{Bondi2003}. In order to take the noise variation across the map into account, a random sample is also created, using pixels between 17 and 42\,arcsec away (in a random direction) from each Gold quasar position. This was to ensure that the new position was well outside the beam covering the Gold candidate position (i.e. more than two beam sizes away), and so avoid any possible correlation in the flux values. Also, using a fixed direction would have introduced a systematic error due to artefacts arising during the radio imaging process, caused by the 6-order symmetry of the VLA's $uv$-coverage. Repeated 1000 times per Gold candidate, this method ensures that all of the random pixels lie within annular loci, centred on a Gold position.

\begin{figure*}
\includegraphics[scale=0.8]{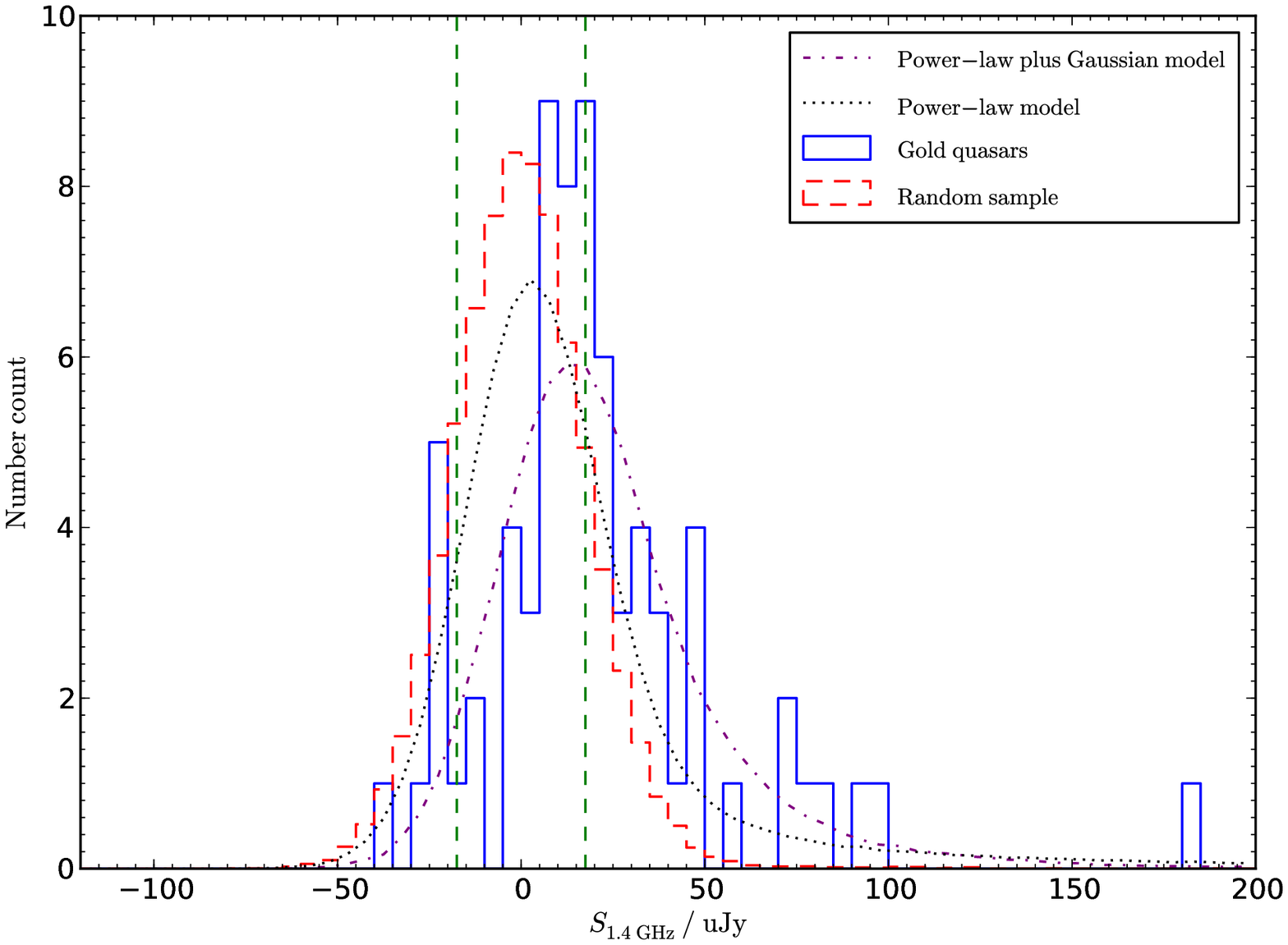}
\caption{Radio flux-density distributions, with the blue (solid) histogram corresponding to the Gold QSOs. The red (dashed) histogram corresponds to the random sample, where the fluxes are extracted from pixels 17 to 42\,arcsec away from the Gold QSO positions. (Their number counts have been scaled to aid comparison.) The green, dashed lines demarcate the mean noise level of the VLA map, at $\pm$17.5\,$\umu$Jy. Two curves represent the distributions for simulated flux densities (see Section~\ref{sec:numbercounts}), if the radio number counts for the quasars is modelled by a power-law (black, dotted line) or a Gaussian-plus-power-law (purple, dash-dotted line). Note that there are two brighter Gold quasars that have radio flux densities outside of the plot range.}
\label{originalKStest}
\end{figure*}

The fraction of negative flux density values is an indication of the level of non-detections, and is significantly lower for the Gold positions than the random positions (Table~\ref{fluxes}). Since the Gold quasar candidates are part of a clean sample, their median radio flux density being higher than that for the random positions is expected. Fig.~\ref{originalKStest} shows flux-density histograms for the quasars and the random positions, and clearly shows a positive tail in the case of the former, reinforcing the evidence for excess radio emission at the positions of the quasars. Within this tail are 10 objects with detections at $>3\sigma$, one of which is above 1\,mJy and appears in both the FIRST and NVSS catalogues \citep{White1997, Condon1998}. Assuming a typical spectral index of $\alpha=-0.7$ (where $S_{\nu} \propto \nu^{\alpha}$), this is the only quasar in the sample that is `radio-loud', according to the definition of $L_{\mathrm{8.4\,GHz}}>10^{25}$\,W\,Hz$^{-1}$ \citep{Hooper1996}. It also satisfies the criteria for radio-loudness defined by \citet{Kellermann1989} and \citet{Miller1990}. However, we reiterate that these are arbitrary boundaries, and although the emission process may be physically different, we continue to include this object in our analysis.

Lastly, although the averages presented are interesting, it must be remembered that stacking leads to a loss of information and it is far more informative to study the entire flux-density distribution (e.g. \citealt{Mitchell-Wynne2014}).

\begin{table*}\centering
\caption{Stacked radio flux densities for the quasar and control samples. `Negative fraction' refers to the number of flux-density values that are negative, and the $p$-value is the result from a Kolmogorov--Smirnov (KS) test. This indicates the probability of the null hypothesis that the two distributions are drawn from the same underlying distribution. The second Gold sample (third row) is created by weighting the appearance of each quasar in accordance with the fraction of its probability distribution function (PDF, Section~\ref{sec:Hewettz}) that survives the $M_{i}\leq-22$ cut imposed in Section~\ref{sec:luminosities}. The last four rows correspond to galaxy samples with stellar mass estimates determined assuming black hole (BH) masses of $10^{8}$\,\Msol\ and $10^{9}$\,\Msol, and specifies whether or not redshift evolution in the $M_{{\rm BH}}$-$M_{*}$ relation was used (see Section~\ref{sec:discussion} for details). In each case, 74 --  the final number of Gold quasars -- is used as the sample size for calculating the uncertainties. We also show the median absolute deviation (MAD), given in brackets, which is less susceptible to outliers in the distribution. Also, the $p$-value quoted is a median, having performed 1000 KS tests with each control sample.}
\begin{tabular}{@{}lrrrr@{}}
 \hline
   Sample& Median flux ($\umu$Jy) & Mean flux ($\umu$Jy) &  Negative fraction & KS test $p$-value\\
 \hline
 Gold quasars & $17.6\pm40.6$  (14.6)&  $57.9\pm32.4$ & 0.19 & N/A\\
 Random positions & $-0.3\pm7.5$ (11.9)& $1.2\pm6.0$ & 0.51&$10^{-12}$\\
Gold quasars weighted by the PDFs& $17.8\pm43.1$ (13.5)&  $63.6\pm34.4$ & 0.18 & N/A\\
$M_{{\rm BH}}=10^{8}$\,\Msol\ and redshift evolution & $2.5\pm21.7$ (12.4)& $7.6\pm17.3$ & 0.44&10$^{-4}$\\
$M_{{\rm BH}}=10^{9}$\,\Msol\ and redshift evolution  & $9.4\pm40.0$ (14.6)& $25.5\pm32.0$ & 0.33&10$^{-2}$\\
$M_{{\rm BH}}=10^{8}$\,\Msol\ and no redshift evolution  & $5.0\pm13.9$ (12.7)& $10.0\pm11.1$ & 0.40&10$^{-3}$\\
$M_{{\rm BH}}=10^{9}$\,\Msol\ and no redshift evolution & $29.9\pm47.0$ (25.4)& $75.3\pm37.5$ & 0.15&10$^{-3}$\\
\hline
\label{fluxes}
\end{tabular}
\end{table*}

\subsection{Statistical detections of radio emission} \label{sec:KStests}

The majority of objects from our quasar sample have radio emission that is below the flux-density limit of the VLA map. Therefore we analyse the radio flux distribution of the quasars to determine whether or not we have a statistical detection of significant radio emission. Fig.~\ref{originalKStest} indicates that we do, but to investigate this quantitatively a two-sample Kolmogorov--Smirnov (KS) test is used. This tests the null hypothesis that the Gold quasar sample and the flux-density measurements at random positions are drawn from the same underlying distribution. We find a KS statistic of $D$ = 0.42 and $p$-value = $10^{-12}$. Therefore the null hypothesis can be rejected, confirming an excess of radio emission for the Gold sample. 
 
\begin{figure*}
\includegraphics[scale=0.73]{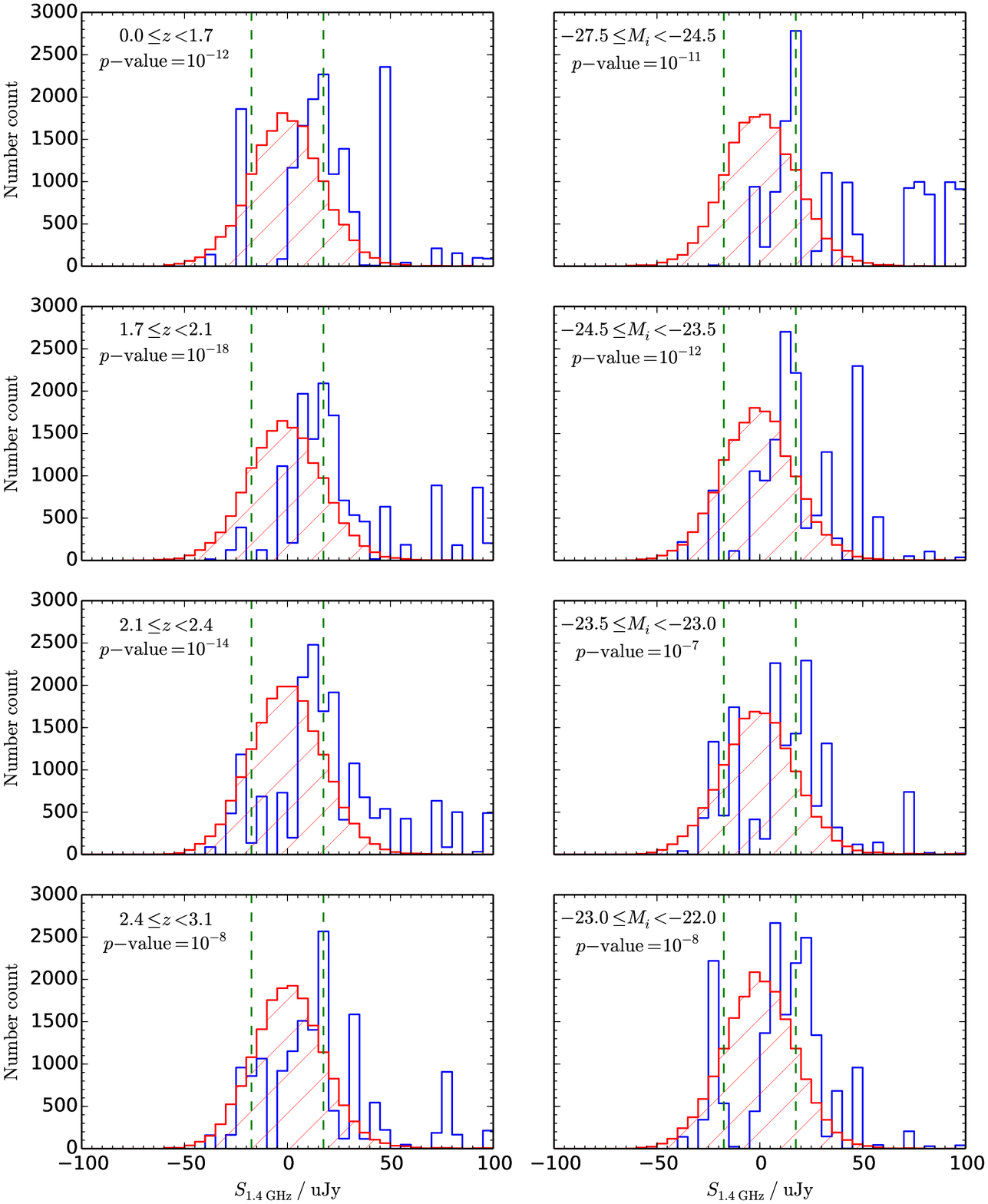}
\caption{The flux-density distribution of Gold quasars (blue histograms), subject to a $M_{i}\leq-22$ cut, binned according to redshift (left column) and absolute {\it i}-band magnitude (right column). To compare with the intrinsic radio noise properties, the flux-density value was also measured at random positions, confined to lie between 17 and 42\,arcsec away from each quasar position (red, hatched histograms, scaled to aid comparison). The dashed green lines are at $\pm$17.5\,$\umu$Jy, the mean rms noise (i.e. 1$\sigma$) level of the VLA map \citep{Bondi2003}. The positive offset of the quasar histogram from the random distribution indicates a statistical detection of faint radio emission. The median $p$-value, resulting from 1000 KS tests between the Gold and random subsamples, is given for each bin (Table~\ref{binnedresults}).}
\label{binnedKStests}
\end{figure*}

We take our analysis a step further by investigating whether the excess radio emission from quasars persists at all redshifts. The 74 quasars in the Gold sample, from Section~\ref{sec:luminosities}, are binned in a probabilistic manner by splitting the simulated objects (derived from the Colour-$z$ PDFs and subjected to a $M_{i}=-22$ magnitude cut) into four redshift ranges. The frequency with which a quasar appears in each bin is recorded, and this is then used to weight the number count distribution for the quasars across the redshift bins. The number counts for the random flux densities, corresponding to a particular quasar as they are (again) constrained to lie between 17 and 42\,arcsec away, are weighted in the same way. To create a Gold subsample for each bin, a number of simulated objects are randomly selected according to how many of the original 74 quasars are sampled within that bin. For the corresponding random subsample, 1000 random positions are selected per Gold object. A KS test is then carried out between the redshift-based subsample and the measurements at random positions, and the procedure repeated 1000 times. The results are shown in Fig.~\ref{binnedKStests} and Table~\ref{binnedresults} and indicate that, for each of the redshift bins, the null hypothesis can be rejected. That is, there is excess radio emission from the quasars at all redshifts. 

\begin{table*}
\centering
\caption{Mean and median radio flux densities for the quasars in our Gold sample, binned by redshift and separately by absolute {\it i}-band magnitude. The number of different quasars appearing in each bin is used to calculate the uncertainties, with the median absolute deviation (MAD) given in brackets. `Negative fraction' refers to the number of flux-density values that are negative. 1000 KS tests are performed per bin, with the median of the resulting $p$-values quoted here.}
\begin{tabular}{@{}crrrr@{}}
 \hline
   Bin range & Median flux ($\umu$Jy) & Mean flux ($\umu$Jy) &  Negative fraction & Median $p$-value\\
 \hline
 $0.0\leq z<1.7$  & $17.3\pm10.1$ (9.8)& $22.5\pm8.0$ & 0.14 & 10$^{-12}$ \\
 $1.7\leq z<2.1$  & $19.9\pm12.6$ (12.0)& $38.8\pm10.0$   & 0.12 & 10$^{-18}$\\
 $2.1\leq z<2.4$  & $18.8\pm15.9$ (13.9)& $34.2\pm12.7$ & 0.19 & 10$^{-14}$\\
 $2.4\leq z<3.1$  & $15.6\pm83.6$ (17.6)& $155.9\pm66.7$ & 0.25 & 10$^{-8}$\\
 $-27.5\leq M_{i}<-24.5$              & $40.8\pm132.2$ (32.9)& $197.6\pm105.5$ & 0.06 & 10$^{-11}$\\
 $-24.5\leq M_{i}<-23.5$              & $19.4\pm12.4$ (14.4)& $34.7\pm9.9$   & 0.14& 10$^{-12}$\\
 $-23.5\leq M_{i}<-23.0$              & $13.3\pm4.1$ (13.8)& $11.8\pm3.2$   & 0.29 & 10$^{-7}$\\
 $-23.0\leq M_{i}<-22.0$              & $14.8\pm5.2$ (8.3)& $13.2\pm4.2$   & 0.21 & 10$^{-8}$\\
\hline
\label{binnedresults}
\end{tabular}
\end{table*}

The Gold quasars were then binned {\it separately} by their absolute {\it i}-band magnitude, using the same probabilistic procedure with regard to the simulated objects. As before, the ranges of these 4 bins were set so that there was roughly the same total number of simulated objects in each. The median $p$-values, calculated from 1000 KS tests between the Gold and random distributions for each $M_{i}$ bin, are also shown in Fig.~\ref{binnedKStests} and in Table~\ref{binnedresults}. We find evidence to reject the null hypothesis, at the $\gg$99.99 per cent confidence level, for each of the bins in absolute $i$-band magnitude.

\begin{figure}
\includegraphics[scale=0.42]{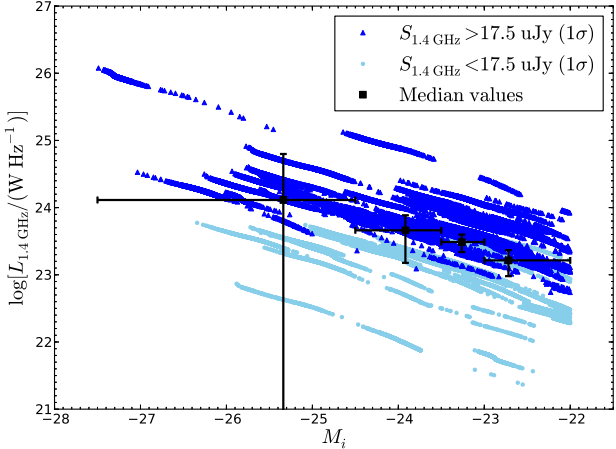}
\caption{Radio luminosity against absolute {\it i}-band magnitude for our quasars. Dark-blue triangles are simulated objects (based on the photometric-redshift probability distributions) with $S_{\mathrm{1.4\,GHz}}>17.5\,\umu$Jy (i.e. $> 1\sigma$), and light-blue circles are those below this radio flux threshold. Overplotted in black are the median luminosities derived for these objects, binned in $M_{i}$. The horizontal error bars indicate the ranges of the $M_{i}$ bins (Table~\ref{binnedresults}), and the ordinate error bars represent the uncertainties on the median radio luminosities.}
\label{radioversusMi}
\end{figure}

Using the individual flux-density measurements and the redshift probability distributions to calculate the distribution in $L_{{\rm 1.4\,GHz}}$ and $M_{i}$, Fig.~\ref{radioversusMi} shows the radio luminosity against the absolute $i$-band magnitude. To perform a correlation test, we randomly select one simulated object per Gold quasar (resulting in a total of 74 simulated objects for each test) and repeat the process 1000 times. A strong correlation is found, with a median coefficient of $r = -0.4\pm0.0$ and median $p$-value = $10^{-4}$. The trend remains when we consider only the objects that have a flux that is above the rms noise level, with $S_{\mathrm{1.4\,GHz}}>17.5\,\umu$Jy (dark-blue triangles in Fig.~\ref{radioversusMi}). The apparent `smearing' of the data points is due to slightly different values of $L_{{\rm 1.4\,GHz}}$ and $M_{i}$ arising from the redshift probability distribution for the quasars (Section~\ref{sec:Hewettz}). If we only use $L_{{\rm 1.4\,GHz}}$ and $M_{i}$ as derived from the best-fit photometric redshift, Colour-$z$, a correlation test for the 74 quasars results in $r = -0.2$ and $p$-value = 0.1. 18 per cent of simulated objects (that survived the $M_{i}<-22$ cut) are missing from Fig.~\ref{radioversusMi}, due to them having a negative flux value extracted from the VLA map. However, the binned median values and associated error bars (in black) {\it do} include quasars with negative fluxes. The resulting negative values of $L_{{\rm 1.4\,GHz}}$ are also used for the correlation test, meaning our sample is not flux-limited in the radio and the trend seen is real.

This correlation between the optical luminosity, which is dominated by thermal emission from the accretion disc, and the non-thermal radio emission has previously been found for radio-loud quasars \citep[e.g.][]{Serjeant1998,Fernandes2011,PunslyZhang2011}, and has been used to infer that the jet-production process is related to the accretion rate, but with black-hole spin playing an important role \citep[e.g.][]{Punsly2011,Tchekhovskoy2011}. However, at the lower radio luminosities that we observe in this sample, we cannot rule out that such a trend is due to the various correlations between black-hole mass, stellar mass and star formation rate \citep[e.g.][]{Noeske2007}.

\subsection{Quasar-related radio source counts} \label{sec:numbercounts}

In this subsection we follow the work of \cite{Kimball2011} and \cite{Condon2013} and investigate the shape of the radio source counts due to quasars. To parameterise the radio background as we probe to lower flux densities, it is necessary to understand the individual contributions from various populations. For example, in the semi-empirical simulation of \citet{Wilman2008, Wilman2010} the number counts of extragalactic objects are broken down into radio-loud and radio-quiet AGN, star-forming galaxies, and starbursts. 

\begin{figure*}
\includegraphics[scale=0.8]{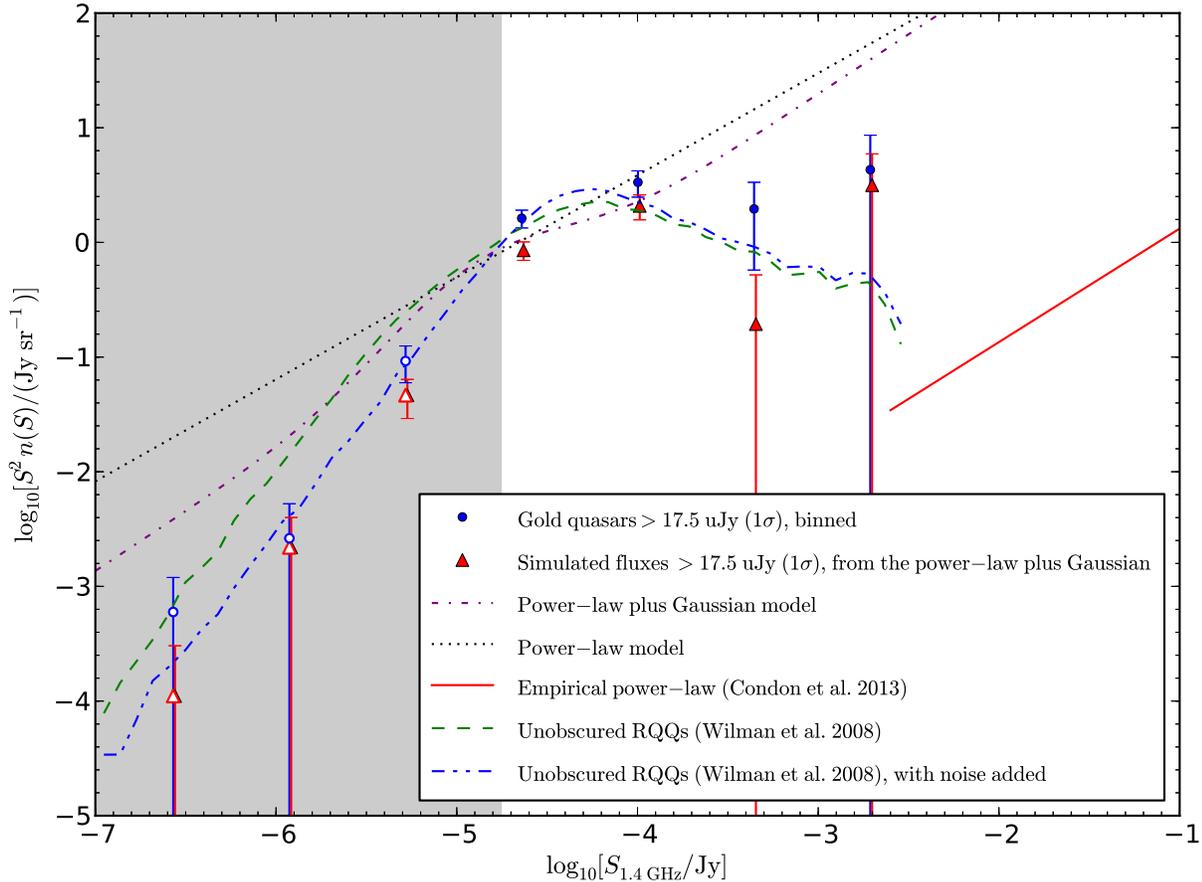}
\caption{The brightness-weighted counts of the Gold quasars (blue dots with error bars), relative to the power-law fit (red, solid line) by \citet{Condon2013}. The equation of their fit is $S^{2}n(S) = (12.9 \pm 1.0)\times S^{0.99\pm0.02}$\,Jy\,sr$^{-1}$. The {\it prescribed} number counts (purple, dash-dotted line) for our simulated points is created by adding a Gaussian to a power-law, as described in Section~\ref{sec:numbercounts}. It does not include the noise contribution later added to create the simulated fluxes (red triangles, slightly offset to show the extent of the error bars). A power-law only model (dotted, black line) is also illustrated. Overplotted are the number counts for RQQs simluated by \citet{Wilman2008}, with and without the addition of noise from the VLA map \citep{Bondi2003}. The shaded region indicates the region fainter than the $1\sigma$ rms flux-density of the radio data, and so the data points corresponding to $S_{{\rm 1.4\,GHz}}< 17.5$\,$\umu$Jy (unfilled symbols) should be treated with caution. Because log-log space is needed to display the data clearly, Gold quasars that have a negative flux density do not appear in the plot. The error bars are Poissonian, depending only on the number of objects appearing in a particular logarithmic bin, $\Delta \log S = 0.643$. (All points are plotted at the centres of the logarithmic bins.)}
\label{numbercounts}
\end{figure*}

In Fig.~\ref{numbercounts} we show the contribution to the radio source counts from our Gold quasar sample using the measurements discussed in Section~\ref{sec:extraction}. The counts are brightness-weighted so that the product, $S^{2}n(S)$, is proportional to the contribution by sources, within each logarithmic bin in flux density, to the sky-background temperature \citep{Condon2012}. Below 1\,mJy we find that a `bump' appears in the number counts, as opposed to a single power-law to lower radio flux densities. Such a bump has also been found in studies based on the SDSS \citep{Kimball2011, Condon2013} and has been attributed to star formation in the quasar host galaxy in both cases. In this subsection we quantify the bump in terms of different parameterisations of the measured radio flux-density distribution, and then explore whether the radio emission is due to star formation or related to the accretion process in Section~\ref{sec:discussion}.
 
To fit the shape of the radio source count contribution from our Gold quasar sample, we need to impose the same survey flux-density limit as our radio data. To begin with, a simulated catalogue is created by drawing fluxes from a function of $S^{2}n(S)$ that follows a particular power-law (see below). Then noise, with rms = 17.5\,$\umu$Jy, is incorporated into the simulated fluxes by adding the flux from randomly-selected pixels of the VLA map used in Section~\ref{sec:extraction}. Effectively we have injected the model fluxes into the radio map and then re-extracted the resultant flux value. This accounts for any small non-uniformity in the noise distribution.

Next, for different combinations of the prescribed slope ($\alpha$) and normalisation ($c$) of the power-law, a chi-squared ($\chi^{2}$) test is performed between the data and the simulated flux distribution. For this we normalise the simulated distribution so that the total number of objects is 74, to match the size of the quasar sample. A minimum in the $\chi^{2}$ value (reduced $\chi^{2}_{\mathrm{min}}=6.2$) is found for the combination of $\alpha = 0.89\pm0.01$ and $c=14000\pm1000$, and the distribution produced by this power-law model is shown by a black, dotted line in Fig.~\ref{originalKStest}. It appears significantly different to the flux-density distribution of the quasars, and this is confirmed by a KS test of the two distributions producing a $p$-value of $7\times10^{-5}$. With no evidence that the simulated fluxes are drawn from the same distribution as the quasar fluxes, we conclude that a power-law is an inadequate description of the number counts. 

We adjust our model by adding a Gaussian contribution to the power-law, and use this to create alternative simulated catalogues. The prescribed function of $S^{2}n(S)$ is then given by
\begin{equation}
\log_{10} [S^{2}n(S)]=\log_{10}[cS^{\alpha}]+A\exp \left( \frac{ \log_{10}[S] - \mu }{ 2 \sigma ^{2} } \right) .
\label{function}
\end{equation}
%where the five parameters governing its shape are: 
%\begin{enumerate}
%\item{$\alpha$ - the slope of the power-law,}
%\item{$c$ - the normalisation of the power-law,}
%\item{$A$ - the amplitude of the Gaussian contribution,}
%\item{$\mu$ - the mean of the Gaussian, and }
%\item{$\sigma$ - the standard deviation of the Gaussian.}
%\end{enumerate} 

Minimising the $\chi^{2}$ value for the data and the resulting simulated flux-density catalogue (reduced $\chi^{2}_{\mathrm{min}}=3.2$), the best-fit model is described by $\alpha = 1.04\pm0.01$, $c = 26000\pm2000$, $A = 0.5\pm0.1$, $\mu = -4.9\pm0.1$, and $\sigma = 0.5\pm0.1$. This gives $p$-value = 0.3 when a KS test is performed between the simulated distribution and the measured radio flux density of the quasars. This indicates that the distribution arising from this power-law plus Gaussian model is indistinguishable from that of the quasars, and can be seen by comparing the purple, dash-dotted line to the blue histogram in Fig.~\ref{originalKStest}. Furthermore, the corresponding contribution to the radio source counts for this model is shown in Fig.~\ref{numbercounts} (red triangles). Note that they do not follow the purple, dash-dotted line there (due to e.g. some objects having negative fluxes, and some objects being shifted between adjacent bins even at the bright end) but do overlap with the real data, thereby illustrating the effect of the noise on the shape of the number counts. This is particularly prominent below $\log_{10}[S_{\mathrm{1.4\,GHz}}]=-4.8$, where the data and the simulated fluxes are within the noise of the VLA radio map (shaded region of Fig.~\ref{numbercounts}). In addition, the relatively small survey area we use (1\,deg$^{2}$) leads to only one Gold quasar in the bins at each end of the flux-density distribution, and the addition of the noise is the dominant reason why these objects are boosted into the brighter flux bins, from the more well-populated regions at fainter flux densities. Indeed, this is why we simulate the flux-distribution based on the model with the noise from the radio map included, to ensure a fair comparison between model and data.

The Gold quasars presented in this paper exceed the SDSS quasar counts by \citet{Condon2013}, as shown by the offset between our data points and their power-law fit (Fig.~\ref{numbercounts}). This is because we include much fainter optically-selected sources, by 6 orders of magnitude in the {\it r}-band, and do so over a redshift range of $0.5 < z < 3.1$ rather than $1.8 < z < 2.5$. Some authors have suggested that, if the radio emission is due to accretion, then the power-law used to describe the brightness-weighted number counts above 1\,mJy should simply extend to lower luminosities. Therefore, any deviation away from this must indicate the presence of radio emission from an alternative process, e.g. star formation (\citealt{Hopkins1998}; \citealt{Kimball2011}). However simulations by \citet{Wilman2008} show that, even when considering the AGN population alone, a distinct `bump' is seen in the number counts. This is because the relation between accretion rate and radio luminosity for radio-quiet quasars does not follow the same scaling relations that are applicable to the radio-loud objects \citep[e.g.][]{Willott1999, Jarvis2001, Fernandes2011}. This may be further accentuated by an emergent star-forming population, and so a combination of the two should be considered.

For our quasar sample, a bump in the number counts is evident below 1\,mJy, and is in agreement with the work of \citet{Condon2013}. As another check with the literature, we perform the same procedure outlined earlier in this subsection using the source counts for radio-quiet quasars as simulated by \citet{Wilman2008}. That is, their model fluxes are injected into the VLA map and then re-extracted, in order for them to have the same noise properties as our binned data and simulated fluxes. Since we cannot impose the same criteria used for our quasar selection, as the simulations do not include reliable absolute magnitudes at optical wavelengths, we simply adjust the normalisation of the resulting number counts (green, dashed line and blue, dash-double-dotted line in Fig.~\ref{numbercounts}), emphasising our interest in the shape alone. Both the Gold quasars (blue dots) and the simulated fluxes (red triangles) incorporate the noise properties of the radio map, and so both datasets should be compared with the blue (dash-double-dotted) line. We find excellent agreement between our data points and the unobscured RQQs modelled by \citet{Wilman2008}, indicated by their overlap within the Poissonian error bars. \cite{Wilman2008} used the empirical relation between X-ray luminosity and radio luminosity for a sample of RQQs, coupled with the X-ray luminosity function to determine the radio luminosity function, and therefore the source counts of the RQQ population. Consequently, such a comparison does not necessarily imply that the radio emission is related to the accretion process, because more luminous AGN are more likely to be powered by more massive black holes that reside in more massive galaxies. These massive galaxies also have a larger gas reservoir from which stars can form, and so generally have higher star formation rates. The strong correlation between stellar mass and star formation rate is in fact observed over multiple decades in stellar mass, and is known as the star formation main sequence \citep{Noeske2007}. `Normal' star-forming galaxies are found to lie along this sequence, whilst starbursts lie above the relation. Therefore, we also further investigate the possibility that the radio emission from the RQQs is due to star formation.

\subsection{Radio emission from AGN or star formation}\label{sec:discussion}

To explore this we use the VIDEO survey to create large samples of quiescent galaxies at similar redshifts to the quasars. This is to determine whether there is excess radio emission from the quasars compared to `normal' galaxies of similar mass. Constructing a sample of galaxies, that is matched in stellar mass to the quasars, is problematic given that the quasar nucleus outshines the host galaxy, thus rendering direct measurements difficult without deep space-based imaging. We therefore adopt an approach whereby we use the observed relation between black-hole mass and stellar mass from \citet{Bennert2011}, $M_{{\rm BH}} \propto M_{*}^{\alpha}(1+z)^{\beta}$ where $\alpha=1.12$ and $\beta = 1.15$. This is quite a moderate evolution compared to other studies \citep[e.g.][]{McLure2006, Peng2006}. We also compare this to the case where we assume no redshift evolution in the relation ($\beta = 0.00$). Rather than calculating the black-hole mass for each quasar, which would require complete spectroscopy that includes either C{\sc iv}$_{1549}$, Mg{\sc ii}$_{2799}$ or H$\beta_{4861}$ broad lines \citep[]{Wandel1999, Vestergaard2002, McLureJarvis2002}, we adopt two fixed black-hole masses of $10^8$ and $10^9$\,\Msol, and calculate the stellar mass of the host accordingly. 

\begin{figure*}
\includegraphics[scale=0.65]{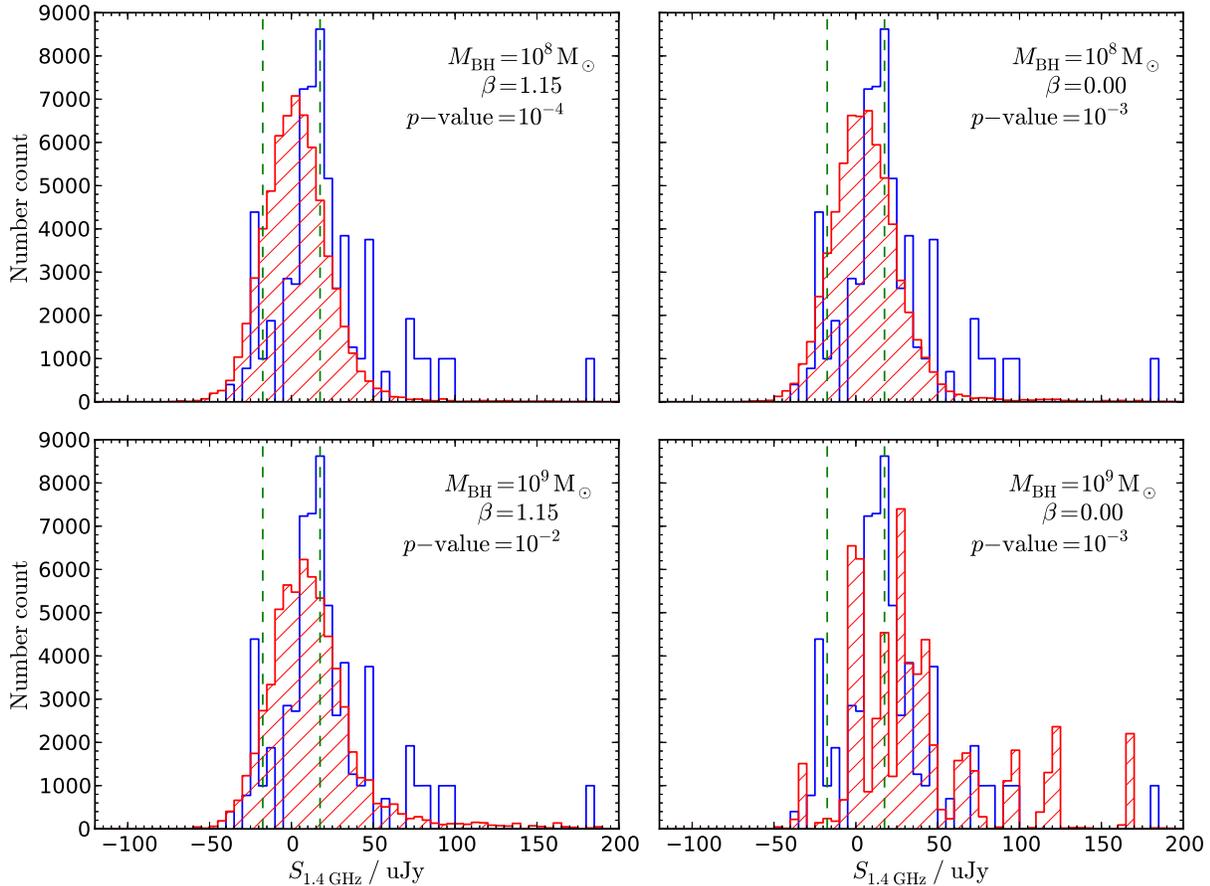}
\caption{Radio flux-density distributions, with the blue histograms corresponding to the Gold quasars. These have been weighted according to the fraction of simulated objects that are brighter than $M_{i}=-22$. (As in Fig.~\ref{originalKStest}, two Gold quasars have radio fluxes outside of this plot range.) The red (hatched) histograms correspond to a control sample, where one galaxy has been matched in $z$ and $M_{*}$ for each simulated object, again having $M_{i}\leq-22$. The stellar masses are estimated assuming that the quasar has $M_{\mathrm{BH}} = 10^{8}$\,\Msol\ (upper panels) or $M_{\mathrm{BH}} = 10^{9}$\,\Msol\ (lower panels). Whether or not redshift evolution in the $M_{{\rm BH}}$--$M_{*}$ relation is used for this calculation is indicated by the value of $\beta$; For evolution we use $\beta=1.15$ (left panels), from the work of \citet{Bennert2011}, whilst $\beta=0.00$ (right panels) corresponds to zero evolution. The median $p$-value, from 1000 KS tests between the Gold and control galaxy samples (Table~\ref{fluxes}), is quoted in the top-right corner of each panel, and the green dashed lines demarcate the mean noise level of the VLA map, at $\pm$17.5\,$\umu$Jy.}
\label{matchedgalaxies}
\end{figure*}

The stellar masses of the normal galaxies are determined by the VIDEO team, using photometric redshifts from {\sc Le PHARE}, in combination with stellar population synthesis models of \cite{BruzualCharlot2003} [see Johnston et al. (in prep.) for more details]. The galaxies are then matched to the quasars in both redshift ($z \pm 0.1$) and stellar mass ($M_{*} \pm 0.3$\,dex) using the full probability distributions in redshift and absolute magnitude, to produce four control samples. For a fair comparison, the Gold quasars are weighted according to the same probability distributions, using the fraction of simulated objects that are brighter than $M_{i}=-22$. With all four control samples, 1000 KS tests are then performed, each time having randomly selected 74 galaxies that are matched to the quasars. Assuming $M_{\mathrm{BH}} = 10^{8}$\,\Msol, multiple KS tests show that the underlying distribution of radio flux density for the quasars and galaxy control sample is significantly different, with median $p=10^{-4}$ and $10^{-3}$, both with and without evolution in the $M_{{\rm BH}}$--$M_{*}$ relation (Table~\ref{fluxes} and Fig.~\ref{matchedgalaxies}). Also, the median flux density of the galaxies is much lower than that of the Gold quasars, for both evolution scenarios. We therefore find evidence for excess radio emission from the quasars, independent of whether or not we assume redshift evolution in the $M_{{\rm BH}}$--$M_{*}$ relation.

Next we look at the control sample created using $M_{\mathrm{BH}} = 10^{9}$\,\Msol\ and $M_{{\rm BH}}$--$M_{*}$ evolution. The median flux density is higher for these matched galaxies compared to those of the first control sample in Table~\ref{fluxes}, as would be expected for galaxies lying on the star-formation main sequence \citep{Noeske2007}. As a result the $p$-value is also larger, indicating that the radio output from these matched galaxies is more significant relative to the quasars' emission. However, the median for this control sample is still below that of the Gold quasar sample, suggesting that emission from the quasars remains in excess. In the case of no redshift evolution in the $M_{{\rm BH}}$--$M_{*}$ relation, the $p$-value shows the quasar and galaxy flux distributions to be distinct, but this time the median flux from the galaxies does exceed that of the Gold quasars. This suggests that, for quasars hosted by the most massive galaxies, star formation could be the cause of the radio emission. 

However, any evolution in the ratio between $M_{{\rm BH}}$ and $M_{*}$ weakens the case for star formation in the quasar hosts. Furthermore, it is unlikely that the majority of our quasars have $10^{9}$\,\Msol\ black-hole masses, given the shape of the black-hole mass function (BHMF) \citep[e.g.][]{McLureDunlop2004, Vestergaard2009}. Additional BHMFs are collated in the review by \citet{Kelly2012}, showing that $10^{8}$\,\Msol\ black holes outnumber $10^{9}$\,\Msol\ black holes by a factor of $\sim$10--20, throughout the redshift range $0<z<3$. Although we could use a realistic black-hole mass distribution for our analysis, rather than a single fixed black-hole mass, we choose not to do so as this would add another layer of complexity. By simply using assumed values, we can more clearly see the impact of the black-hole mass when investigating the origin of the radio emission in RQQs. In addition, most work in the literature uses data from the SDSS to study the BHMF, rather than quasars selected to the same depth as our sample. Therefore we would expect our quasars to have, on average, lower-mass black holes.

Star formation in the host galaxies of quasars has also recently been studied by several authors. \cite{Bonfield2011} used a sample of SDSS quasars and data from the {\em Herschel}--ATLAS \citep{Eales2010} to show that the star formation is correlated to the quasar accretion luminosity as well as the redshift. In a similar study, \citet{Rosario2013} find that the star formation properties of quasars are consistent with a model where the host galaxy lies on the star formation main sequence (\citealt{Noeske2007}; \citealt{Whitaker2012}). These are both based on the assumption that the AGN emission dominates the optical light and does not make a strong contribution to the far-infrared, which is used to calculate the star formation rate. In addition, support for radio emission from quasars being related to star formation is provided by recent results on sub-mJy radio source populations \citep{Padovani2014}, although we note that the AGN in the study of Padovani et al. are generally much fainter optically than the quasars considered here.

\begin{figure*}
\includegraphics[scale=0.45]{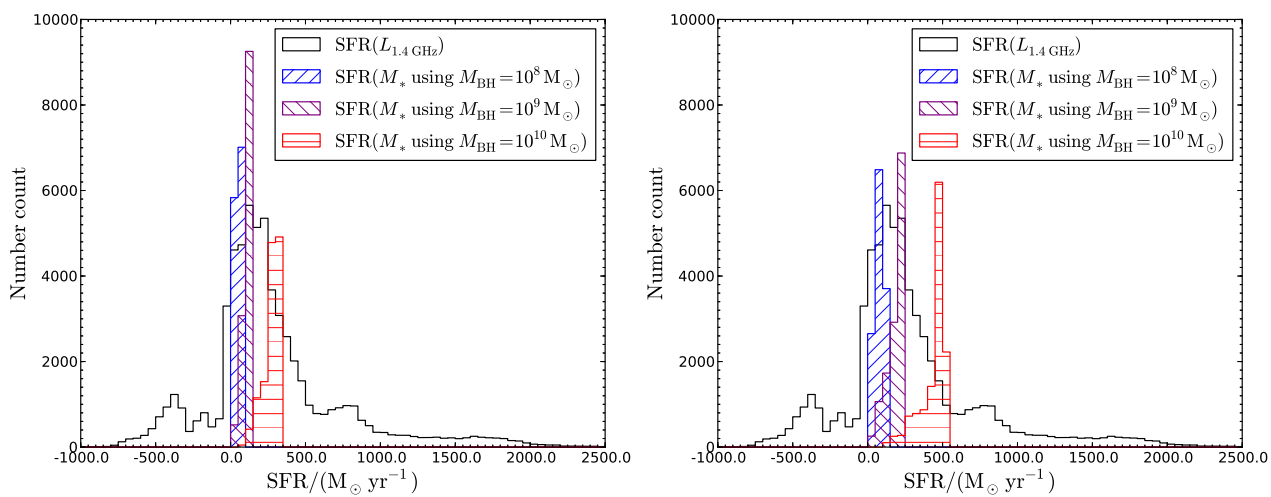}
\caption{Distributions in the SFR for the quasars, as derived from either the radio luminosity (black histograms) or an estimate of the stellar mass. The latter is calculated using a combination of assumed black-hole mass -- $10^{8}$\,\Msol\ (blue histograms), $10^{9}$\,\Msol\ (purple histograms) or $10^{10}$\,\Msol\ (red histograms) -- and evolution (left) or no evolution (right) in the $M_{{\rm BH}}$--$M_{*}$ relation. These hatched histograms have been scaled by a factor of 0.2 to ease comparison with the SFR($L_{{\rm 1.4\,GHz}}$) distribution. Note that two quasars, brightest in radio luminosity, lie beyond the range of these plots.}
\label{SFRhistograms}
\end{figure*}

\begin{figure*}
\includegraphics[scale=0.45]{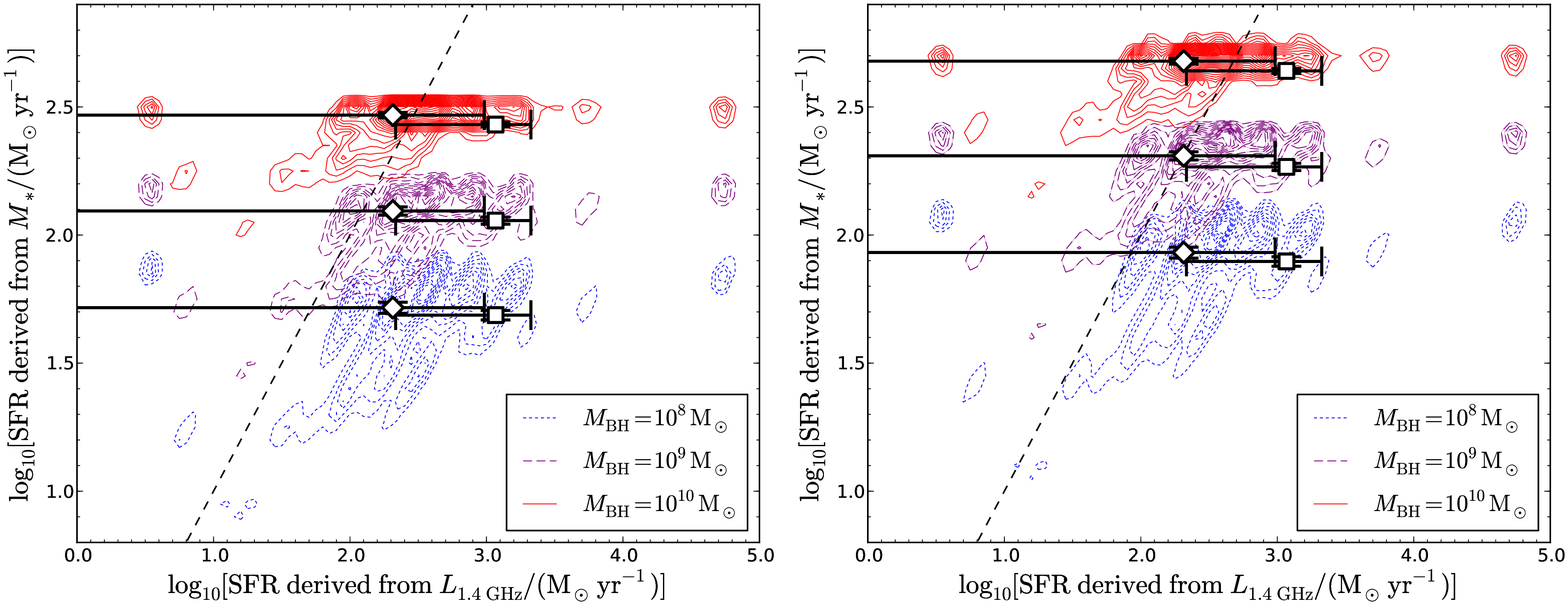}
\caption{The SFR calculated from an estimate of the stellar mass, derived for different black-hole masses (see legend), against the SFR calculated from the extracted radio flux density at 1.4\,GHz. The stellar masses used are calculated {\it with} (left) and {\it without} (right) evolution in the $M_{{\rm BH}}$--$M_{*}$ relation. The simulated quasars (derived from the photometric-redshift probability distributions) are represented by blue (dotted), purple (dashed) and red (solid) contours (interval = 50), for assumed black-hole masses of $10^{8}$\,\Msol, $10^{9}$\,\Msol\ and $10^{10}$\,\Msol, respectively. The dashed line indicates the one-to-one relation, if star formation (consistent with the SF-mass sequence) accounts for all of the radio emission. Quasars that have a measured negative radio flux are not shown, however they do contribute to the median and mean SFRs (white diamonds and squares, respectively, with error bars giving the uncertainties - see Table~\ref{SFRtable}).} 
\label{TestingSFmasssequence}
\end{figure*}

We therefore further investigate the origin of the radio emission in our quasar sample by comparing two independent estimates of the star formation rate. The first value uses the 1.4\,GHz flux extracted from the VLA map, and uses the relation between radio luminosity and SFR, as quantified by \citet{Yun2001} for a sample of IR-selected galaxies. A second estimate of the SFR is based on the assumption that the quasar host galaxies lie on the star formation main sequence, as found by e.g. \citet{Karouzos2014}. Of course this may not be the case for our quasars, whose radio emission may instead be due to starbursts, but it is a suitable first approximation. For this we use the redshifts (simulated from the photometric-redshift probability distributions) and typical black-hole masses for a quasar ($10^8$--$10^{10}$\,\Msol) to estimate the total stellar mass, as previously described. (An insufficient number of galaxies in the VIDEO catalogue could be matched to the quasars when assuming $M_{{\rm BH}}=10^{10}$\,\Msol, and so this control sample was not created.) These stellar mass estimates are then combined with the finding that SFR $\propto M_{*}^{0.6}$ \citep{Whitaker2012}, which describes the redshift evolution of the star formation main sequence. The distribution in the SFRs are presented in Fig.~\ref{SFRhistograms}, with the mean and median values for each distribution provided in Table~\ref{SFRtable}. To further aid comparison, these SFRs are represented by contours in Fig.~\ref{TestingSFmasssequence}, effectively acting as probability distribution functions for a particular black-hole mass. 

\begin{table*}\centering
\caption{Mean and median star formation rates (SFRs) for the simulated objects, according to how their SFR was derived. The first row corresponds to the relation of \citet{Yun2001} being applied to the extracted radio fluxes. The remaining calculations of SFR involve stellar mass estimates, for a particular assumed black hole (BH) mass with (or without) $M_{{\rm BH}}$-$M_{*}$ evolution (see Section~\ref{sec:discussion} for details). In each case, 74 --  the final number of Gold quasars -- is used as the sample size for calculating the uncertainties, with the median absolute deviation (MAD) given in brackets.}
\begin{tabular}{@{}lrr@{}}
 \hline
  Estimate used to derive the SFR& Median SFR (\Msol\ yr$^{-1}$) & Mean SFR (\Msol\ yr$^{-1}$) \\
 \hline
$L_{\mathrm{1.4\,GHz}}$ & $206.0\pm950.1$ (187.9)&  $1165.9\pm758.2$ \\
$M_{*}$ using $M_{{\rm BH}}=10^{8}$\,\Msol\ and redshift evolution & $52.1\pm2.6$ (12.5)& $48.7\pm2.1$ \\
$M_{*}$ using $M_{{\rm BH}}=10^{9}$\,\Msol\ and redshift evolution  & $124.2\pm4.6$ (18.2)& $113.9\pm3.6$ \\
$M_{*}$ using $M_{{\rm BH}}=10^{10}$\,\Msol\ and redshift evolution  & $293.9\pm7.3$ (14.0)& $269.9\pm5.8$ \\
$M_{*}$ using $M_{{\rm BH}}=10^{8}$\,\Msol\ and no redshift evolution  & $85.5\pm4.2$ (19.9)& $78.9\pm3.4$ \\
$M_{*}$ using $M_{{\rm BH}}=10^{9}$\,\Msol\ and no redshift evolution & $203.8\pm7.5$ (27.2)& $184.5\pm6.0$ \\
$M_{*}$ using $M_{{\rm BH}}=10^{10}$\,\Msol\ and no redshift evolution & $477.2\pm12.5$ (23.5)& $437.1\pm10.0$ \\
\hline
\label{SFRtable}
\end{tabular}
\end{table*}

We see that there is a mismatch between the two sets of star formation rates calculated for the simulated objects associated with the final 74 quasars (Fig.~\ref{TestingSFmasssequence}), independent of whether the $M_{{\rm BH}}$--$M_{*}$ relation is evolving or not. Quasars are most likely to have black-hole masses of $10^8 < M_{\rm BH} < 10^9$\,\Msol\ \citep[e.g.][]{McLureDunlop2004}, but the SFR implied by their stellar mass indicates that this is an order of magnitude lower than expected from the radio emission. Indeed, if there is $M_{{\rm BH}}$--$M_{*}$ evolution, the quasars would need to have $M_{\mathrm{BH}} \approx10^{10}$\,\Msol\, in order for their SFRs to have a one-to-one correspondence (lying on the dashed line). Assuming no evolution, the black-hole mass would need to be $>10^9$\,\Msol\ if we are to explain the total radio emission as due to star formation alone. Note that only objects with positive radio fluxes appear in Fig.~\ref{TestingSFmasssequence}, which may introduce a slight bias. However, this is taken into account by the median and mean SFRs that are also shown, as all objects were used for their calculation. 

Therefore, either our assumption that the quasar hosts lie on the SF-mass sequence is incorrect, or the AGN is contributing a significant fraction towards the radio flux. We argue that the latter is the most probable explanation because, based on work using far-infrared data \citep{Bonfield2011, Rosario2013}, it is unlikely that the quasar host galaxies are undergoing massive starburst activity. 

As mentioned earlier, the selection criteria we use to create our sample mean that we are biased towards the brightest quasars (with high accretion rates/high black-hole masses) and the faintest host galaxies (possibly with low stellar mass). Since these objects are best-fit by a Type-1 quasar template, and dust has less of an effect in the $K_{\rm S}$-band, we do not believe that obscuration is the reason for the faintness of the host galaxy in the optical and near-infrared bands. Instead, these objects may have low stellar mass and therefore low star formation rates, and so we would expect there to be less radio emission due to star formation as a result of our selection.

On a similar note, it may seem inappropriate to apply the standard $M_{{\rm BH}}$--$M_{*}$ relation, given the biases described above and our selection of only 20 per cent of the total quasar population (see end of Section~\ref{sec:Hewettmodel}). By doing so, for each assumed black-hole mass, we may in fact be overestimating the stellar mass. Consequently, the SFR of the host is also overestimated, and the true proportion of radio emission due to accretion would be even greater. \citet{Kimball2011} and \citet{Condon2013} use SDSS-selected quasars, which are also biased towards high accretion and low galaxy mass, albeit not to the same degree as our sample. (This is because our selection is strongly dependent on the $K_{\rm S}$-band, where the host galaxy and quasar emission are generally more comparable, relative to selection at optical wavelengths.) We also note however, that if the relation between optical luminosity and radio luminosity shown in Fig.~\ref{radioversusMi} is due to the accretion process then we would expect a greater degree of AGN-related radio emission in the SDSS samples. Therefore carrying out the same analysis for their quasars would also lead to an overprediction of the amount of radio emission due to star formation. Thus our conclusions, that the accretion process is the primary origin of the radio emission at $S_{1.4\rm\,GHz} < 1$\,mJy, are strengthened.

\section{Conclusions} \label{sec:conclusions}

As we probe to lower radio luminosities, it is vital that we fully appreciate the various populations contributing to the radio background. Of particular importance are star-forming galaxies and radio-quiet quasars (RQQs). These two types of sources may be confused for one another, owing to the similar amount of low-level radio emission they produce. They may even be composite, having both ongoing star formation and black-hole accretion. To disentangle the two, and understand their roles in galaxy evolution, these objects must be better studied.

\begin{enumerate}
\renewcommand{\theenumi}{(\arabic{enumi})}

\item{In this paper we have described a robust method for the selection of RQQs in the VIDEO Survey \citep{Jarvis2013}. By applying a morphology cut and combining AGN template fitting with photometry, a region within {\it gJK$_{\rm S}$} colour space is defined that successfully selects Type-1 (i.e. unobscured) quasars. These make up our `Gold' candidate quasar sample, which (following an absolute $i$-band magnitude cut, $M_{i}\leq-22$) comprises 74 quasars.}\\

\item{We use optical spectroscopy from both the SALT and VVDS, which provides accurate redshifts for 26 per cent of the Gold objects. Our photometric redshifts are shown to be reliable, with $\sigma_{\Delta z/(1+z)} = 0.046$ when calculated across the quasars with spectroscopic redshifts.}\\

\item{Using a probabilistic method to bin the Gold quasars by redshift, and separately by absolute {\it i}-band magnitude, two-sample Kolmogorov--Smirnov (KS) tests were performed to compare the radio flux-density distribution of the quasars with that of random positions. The results show that there is an excess of radio emission, at the $\gg$99.99\% confidence level, for quasars belonging to each of the redshift and luminosity bins. We also find a trend across the $M_{i}$ bins, with higher radio luminosity correlating with increased optical luminosity. This provides indirect evidence for a relation between the accretion process and the source of the radio emission.}\\

\item{By comparing the radio flux-density distribution for the Gold quasars with that for control galaxies matched in redshift and stellar mass, we find that the quasars have excess radio flux when assuming the most reasonable values of black-hole mass ($\sim 10^8$\,\Msol), independent of whether we adopt an evolution in the black-hole mass -- stellar mass relation. Star formation could explain the radio emission for quasars with the most massive ($\sim 10^{9}$\,\Msol) black holes. However, this is only if we assume that there is {\it no} evolution in the $M_{{\rm BH}}$--$M_{*}$ relation. This indicates that accretion is the primary origin of the quasars' total radio emission.}\\

\item{The contribution to radio source counts from quasars cannot be described by a power-law alone, due to a `bump' appearing below 1\,mJy. We find that a power-law plus Gaussian model for the source count distribution reproduces the behaviour over the range $-4.8<\log S_{{\rm 1.4\,GHz}}<-2.5$. We also find that such a source count distribution is in good agreement with the source count model for unobscured RQQs, as simulated by \citet{Wilman2008}. The appearance of this feature is consistent with previous work in the literature (e.g. \citealt{Kimball2011}; \citealt{Condon2013}). These authors suggest that a star-forming population is beginning to dominate over the AGN population at low radio luminosities. We make two independent calculations of SFR, one based on the expected stellar mass of the quasar's host galaxy, and the other using the radio luminosity. A comparison of the two indicates that the AGN in these RQQs are making the dominant contribution to the total radio emission. Although we cannot rule out the possibility that the host galaxies are undergoing starburst activity, we note that studies using far-infrared data from {\em Herschel} appear to rule out such prodigious star formation in quasar hosts.}\\ 

Over the next decade studies of the radio properties of radio-quiet AGN will move from the statistical work such as described in this paper, to direct detections with the new generation of radio telescopes. For example, the MeerKAT--MIGHTEE continuum survey \citep{Jarvis2012} will survey 35\,deg$^2$ to 1\,$\umu$Jy rms, and ASKAP--EMU \citep{Norris2011} will image the whole sky to 10\,$\umu$Jy rms, thus providing direct measurements of the contribution of radio-quiet quasars to the total radio source counts. 

\end{enumerate}

%\vspace{-0.65cm}
\section*{Dedication}
This paper is dedicated to the memory of Steve Rawlings, who was an absolute pleasure to work with. 

\section*{Acknowledgements}

SVW would like to thank L. Miller, I. Heywood, D. G. Bonfield, and D. J. B. Smith for their discussions over the course of this work. In addition, P.C. Hewett, whose useful suggestions helped to strengthen the paper. The first author also wishes to acknowledge support provided through an STFC studentship. 

This work is based on data products produced by the Cambridge Astronomy Survey Unit on behalf of the VIDEO Consortium. The observations for these products were made with ESO Telescopes at the La Silla Paranal Observatory under programme ID 179.A-2006 (PI: Jarvis).

We also use observations obtained with MegaPrime/MegaCam, a joint project of CFHT and CEA/DAPNIA, at the Canada-France-Hawaii Telescope (CFHT) which is operated by the National Research Council (NRC) of Canada, the Institut National des Science de l'Univers of the Centre National de la Recherche Scientifique (CNRS) of France, and the University of Hawaii. This work is based in part on data products produced at TERAPIX and the Canadian Astronomy Data Centre as part of the Canada-France-Hawaii Telescope Legacy Survey, a collaborative project of NRC and CNRS.

Additional observations reported in this paper were obtained with the Southern African Large Telescope (SALT), under proposal code 2012-2-RSA\_OTH-005 (PI: Jarvis).

{\sc IRAF} is distributed by the National Optical Astronomy Observatory, which is operated by the Association of Universities for Research in Astronomy (AURA) under cooperative agreement with the National Science Foundation.

%\vspace{-0.15cm}
\bibliographystyle{mn2e_mod}
\bibliography{SarahWhite_RQQs_MNRAS_accepted}
%\bibdata{MasterReferences}

%\vspace{-0.25cm}
\appendix

\section[]{Magnitude conversions} \label{sec:conversions}

Table~\ref{offsets} gives the values used to convert between the AB and Vega magnitude systems, i.e. $\mathrm{mag}_{\mathrm{Vega}} = \mathrm{mag}_{\mathrm{AB}} + \mathrm{Vega\ offset}$. The $Z$ band does not appear in the table as it was not used for the SED-fitting (Section~\ref{sec:LePHARE}).

\begin{table}
\centering
\caption{AB to Vega magnitude conversions for the VIDEO--CFHTLS--D1 data.}
\begin{tabular}{@{}cccccc@{}}
 \hline
 Band & $\lambda$ &  Vega & Band & $\lambda$ &  Vega\\
     &   ($\umu$m) &  offset &     &   ($\umu$m) &  offset \\
 \hline
 {\it u} & 0.38 & -0.434 &  {\it Y} & 1.02 & -0.614 \\
 {\it g} & 0.48& 0.078 &  {\it J} & 1.25  & -0.929 \\
 {\it r} & 0.63 & -0.165 &  {\it H} & 1.65 & -1.375 \\
 {\it i} & 0.77 & -0.402 &  {\it K}$_{\rm S}$ & 2.15 & -1.836\\
 {\it z} & 0.89 & -0.536 & \\

\hline
\label{offsets}
\end{tabular}
\end{table}

\section[]{Morphology classification} \label{sec:completeness}

The parameter K\_CLASS\_STAR, an output of {\sc SExtractor}, is used as an indication of how point-like an object appears. To test the use of K\_CLASS\_STAR $>$ 0.9 as a criterion for selecting quasars (Section~\ref{sec:Kband}), simulations of AGN-plus-host-galaxies are created using the parameters detailed in Table~\ref{pointlikesimulations}. (Also note that a S{\' e}rsic index of 4 is used for simulating the hosts.) The results presented in Table~\ref{completeness} indicate the level of completeness, for different ratios of AGN flux ($f_{\rm AGN}$) to total flux ($f_{\rm total}$), when this morphological cut is made. To clarify, $f_{\rm total} = f_{\rm AGN} + f_{\rm galaxy}$. 

\begin{table}
\centering
\caption{The parameter values used to simulate AGN and their host galaxies. The magnitudes of the AGN and the host are constrained by the assigned combination of total magnitude, $K_{{\rm S}}$, and flux ratio, $f_{\rm AGN}/f_{\rm total}$.}
\begin{tabular}{@{}lccr@{}}
 \hline
Parameter & Min. & Max. &Distribution\\
 \hline
$K_{\rm S}$ magnitude & 18.0 & 23.5& In steps of 0.5\\
$f_{\rm AGN}/f_{\rm total}$ & 0.0&1.0 & In steps of 0.1\\
$r_{\rm eff}$/arcsec & 0.3 &2.1&In steps of 0.3\\
\hline
\label{pointlikesimulations}
\end{tabular}
\end{table}

\begin{table}
\caption{The fraction of simulated objects having K\_CLASS\_STAR $>$ 0.9, for a given combination of AGN flux ($f_{\rm AGN}$) and total flux ($f_{\rm total}$), as a function of total magnitude, $K_{\rm S}$.}
\begin{tabular}{ccccccc}
 \hline

 & \multicolumn{6}{c}{Total magnitude, $K_{\rm S}$}   \\ 
  $f_{\rm AGN}/f_{\rm total}$& {\bf 18.0} & {\bf 18.5} & {\bf 19.0} & {\bf 19.5} & {\bf 20.0} & {\bf 20.5}   \\
\hline
{\bf 0.0} & 0.00 & 0.00 & 0.00 & 0.00 & 0.00 & 0.00  \\ %magAGN-magGalaxy=12.5
{\bf 0.1} & 0.00 & 0.00 & 0.01 & 0.03 & 0.02 & 0.06  \\ %magAGN-magGalaxy=2.39
{\bf 0.2} & 0.00 & 0.09 & 0.10 & 0.14 & 0.09 & 0.12  \\ %magAGN-magGalaxy=1.51
{\bf 0.3} & 0.03 & 0.25 & 0.36 & 0.31 & 0.31 & 0.36  \\ %magAGN-magGalaxy=0.92
{\bf 0.4} & 0.75 & 0.85 & 0.98 & 0.94 & 0.82 & 0.85  \\ %magAGN-magGalaxy=0.44
{\bf 0.5} & 0.86 & 0.88 & 1.00 & 1.00 & 1.00 & 1.00  \\ %magAGN-magGalaxy=0.00
{\bf 0.6} & 0.68 & 0.83 & 1.00 & 1.00 & 1.00 & 1.00  \\ %magAGN-magGalaxy=-0.44
{\bf 0.7} & 0.07 & 0.24 & 1.00 & 1.00 & 1.00 & 1.00  \\ %magAGN-magGalaxy=-0.92
{\bf 0.8} & 0.21 & 0.99 & 1.00 & 1.00 & 1.00 & 1.00  \\ %magAGN-magGalaxy=-1.51
{\bf 0.9} & 1.00 & 1.00 & 1.00 & 1.00 & 1.00 & 1.00  \\ %magAGN-magGalaxy=-2.39
{\bf 1.0} & 1.00 & 1.00 & 1.00 & 1.00 & 1.00 & 1.00  \\ %magAGN-magGalaxy=-12.5
 \hline
 & \multicolumn{6}{c}{Total magnitude, $K_{\rm S}$}   \\ 
  $f_{\rm AGN}/f_{\rm total}$&  {\bf 21.0} & {\bf 21.5} & {\bf 22.0} & {\bf 22.5} & {\bf 23.0} & {\bf 23.5}   \\
\hline
{\bf 0.0} & 0.00 & 0.00 & 0.01 & 0.00 & 0.00 & 0.00  \\ %magAGN-magGalaxy=12.5
{\bf 0.1} & 0.03 & 0.04 & 0.04 & 0.00 & 0.00 & 0.00   \\ %magAGN-magGalaxy=2.39
{\bf 0.2} & 0.15 & 0.13 & 0.06 & 0.00 & 0.00 & 0.00  \\ %magAGN-magGalaxy=1.51
{\bf 0.3} & 0.49 & 0.35 & 0.04 & 0.00 & 0.00 & 0.00   \\ %magAGN-magGalaxy=0.92
{\bf 0.4} & 0.83 & 0.69 & 0.20 & 0.00 & 0.00 & 0.00   \\ %magAGN-magGalaxy=0.44
{\bf 0.5} & 0.98 & 0.88 & 0.37 & 0.00 & 0.00 & 0.00   \\ %magAGN-magGalaxy=0.00
{\bf 0.6} & 1.00 & 0.99 & 0.64 & 0.00 & 0.00 & 0.00   \\ %magAGN-magGalaxy=-0.44
{\bf 0.7} & 1.00 & 1.00 & 0.91 & 0.00 & 0.00 & 0.00   \\ %magAGN-magGalaxy=-0.92
{\bf 0.8} & 1.00 & 1.00 & 0.95 & 0.02 & 0.00 & 0.00   \\ %magAGN-magGalaxy=-1.51
{\bf 0.9} & 1.00 & 1.00 & 0.99 & 0.13 & 0.00 & 0.00   \\ %magAGN-magGalaxy=-2.39
{\bf 1.0} & 1.00 & 1.00 & 1.00 & 0.17 & 0.00 & 0.00  \\ %magAGN-magGalaxy=-12.5
\hline
\label{completeness}
\end{tabular}
\end{table}

\section[]{Colour selection of quasars} \label{sec:equations}

The following straight-line equations define the `Gold' selection region of Fig.~\ref{all}, where $x = J-K$, $y = g-J$. To reiterate, the magnitudes used are AB. To convert to the Vega magnitude system, see Table~\ref{offsets}.
\begin{align}
x = -0.30\\
y=-1.00 \hspace{1.2cm} -0.30<x<1.10\\ 
y=0.65 \hspace{1.2cm} -0.30<x<1.10\\ 
x=1.10
\end{align}
This allows the reader to use the same selection method if they wish to compare quasar samples.

\section[]{SALT spectra} \label{sec:allspectra}

\begin{figure*}
\includegraphics[scale=0.73]{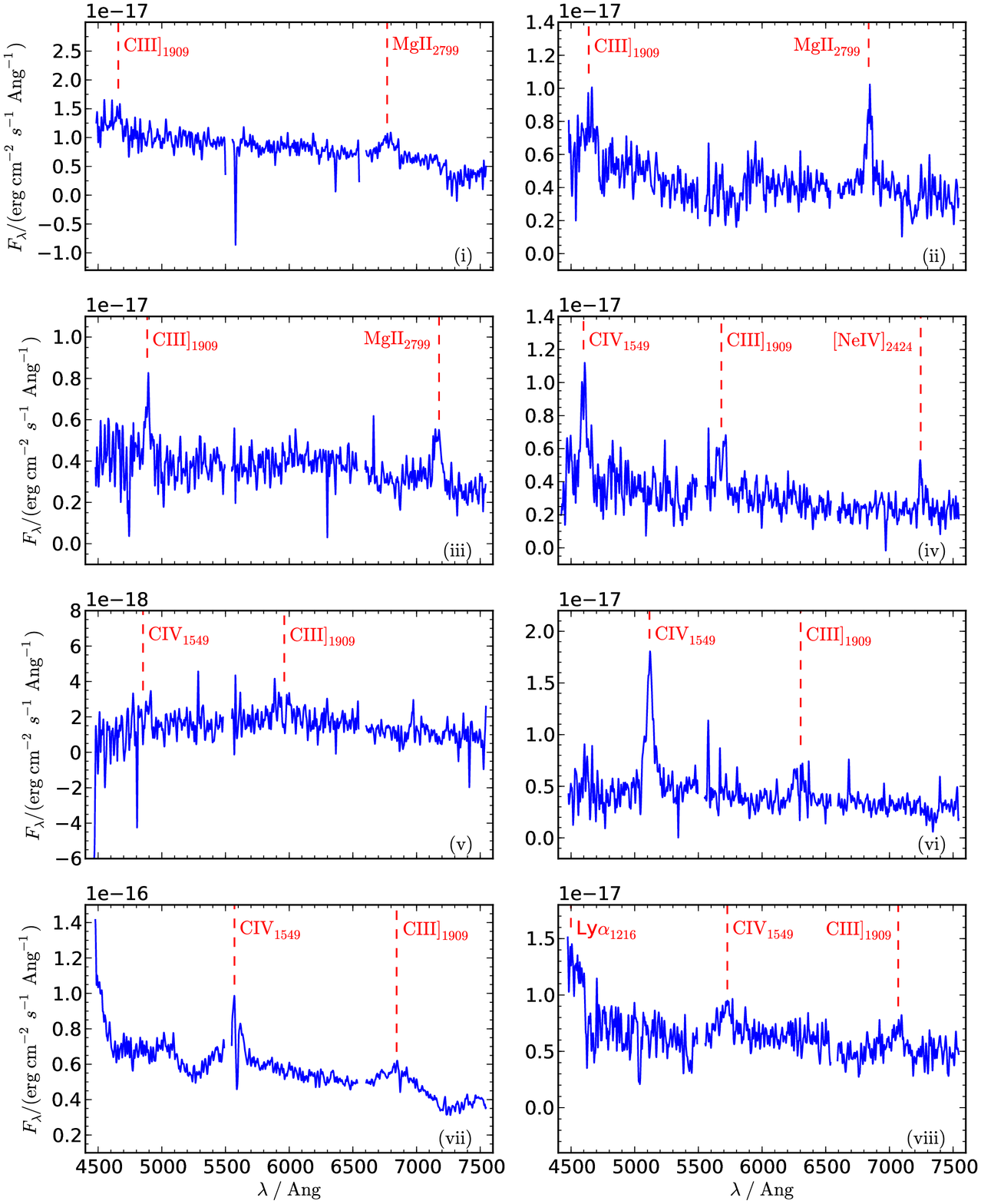}
\caption{SALT spectra for 8 Gold objects: (i) A quasar at $z$ = 1.4 (RA = 02:26:52.14, Dec = -04:05:57.1). (ii) A quasar at $z$ = 1.4 (RA = 02:25:52.16, Dec = -04:05:16.1). (iii) A quasar at $z$ = 1.6 (RA = 02:26:39.83, Dec = -04:20:04.4). (iv) A quasar at $z$ = 2.0 (RA = 02:26:18.59, Dec = -04:11:01.1). (v) A quasar at $z$ = 2.1 (RA = 02:26:33.31, Dec = -04:29:47.8). (vi) A quasar at $z$ = 2.3 (RA = 02:26:12.64, Dec = -04:34:01.4). (vii) A quasar at $z$ = 2.6 (RA = 02:27:40.55, Dec = -04:02:51.1). There is possible Ly$\alpha$ emission at the spectrum's blue end, whilst peaks at $\lambda=4842, 6783, 7251$\,\AA\ are due to cosmic rays. This is the brightest object in the Gold sample, and the only one to appear in both FIRST and NVSS radio catalogues. (viii) A quasar at $z$ = 2.7 (RA = 02:27:09.03, Dec = -04:55:10.1).}
\label{allspectra}
\end{figure*}

Fig.~\ref{allspectra} shows spectra taken with the Southern African Large Telescope (SALT), confirming 8 Gold candidates to be quasars. These have been smoothed with a Gaussian window that covers 20 pixels. (Chip gaps are over the ranges $5500\lesssim \lambda ({\rm \AA}) \lesssim 5545$ and $6550\lesssim \lambda ({\rm \AA}) \lesssim 6585$.) The resolving power, $R=\lambda / \Delta \lambda \approx 600$, gives a resolution of 10\,\AA. Prominent emission lines are labelled, except in panel (v) where the two labels indicate resonant absorption features.

\section[]{Details of the final 74 quasars} \label{sec:74quasars}

Spread over two pages is a table presenting the optical and near-infrared photometry of the final 74 quasars, as well as their radio flux density and redshift (Table~\ref{74quasars}).

\bsp

\begin{landscape}
\begin{table}

\caption{Positions, AB magnitudes, radio flux densities, and redshifts for the final 74 Gold quasars. $z_{spec}$ is the SALT or VVDS redshift, where available. \newline * This VVDS object is thought to have been assigned an incorrect redshift. The corrected value is given here.}
\begin{tabular}{@{}ccccccccccccrr@{}}
 \hline
R.A. (hms)& Dec. (dms) & $u$ & $g$ & $r$ & $i$ & $z$ & $Z$ & $Y$ & $J$ & $H$ & $K_{{\rm S}}$ & $S_{{\rm 1.4\,GHz}}/\umu$Jy & $z_{spec}$ \\
 \hline

    02:24:07.43 &    -04:45:55.5 &    23.08 &    22.83 &    22.67 &    22.34 &    22.27 &    22.52  &    22.06 &     22.42 &     22.27 &     21.88         &39.69&        -           \\       % VIDEO ID =     1034623    
  02:24:09.79 &    -04:34:04.9 &    22.20  &    21.93 &    21.57 &    21.59 &    21.57 &    22.32  &    21.82 &     21.81 &     21.52 &     21.16   &-37.39&         -            \\       % VIDEO ID =     1062537    
  02:24:09.81 &    -04:29:48.7 &    22.05 &    21.67 &    21.59 &    21.58 &    21.43 &    21.31  &    21.49 &     21.44 &     21.17 &     20.89  &-24.96&         -           \\       % VIDEO ID =     1073625    
  02:24:09.89 &    -04:47:18.1 &    20.59 &    20.19 &    20.07 &    19.72 &    19.74 &    19.94  &    19.87 &     19.89 &     19.96 &     19.81   &180.21&          -            \\       % VIDEO ID =     1031287    
  02:24:13.47 &    -04:52:10.4 &    21.80  &    21.09 &    21.13 &    21.06 &    20.89 &    21.01  &    20.66 &     20.65 &     20.59 &     20.40   &-1.92&         -            \\       % VIDEO ID =     1019065    
  02:24:14.93 &    -04:43:53.7 &    22.85 &    22.41 &    22.38 &    22.05 &    22.10  &    21.91  &    22.18 &     22.07 &     21.69 &     21.67        &-21.51&          -           \\       % VIDEO ID =     1039761    
  02:24:15.30 &    -04:29:55.9 &    22.99 &    22.71 &    22.50  &    22.38 &    22.13 &    22.36  &    22.21 &     22.01 &     21.55 &     20.96         &21.27&       -            \\       % VIDEO ID =     1073415    
  02:24:19.75 &    -04:03:15.7 &    22.38 &    22.12 &    22.21 &    21.89 &    21.61 &    21.80   &    21.82 &     21.81 &     21.61 &     21.03  &20.92&          -            \\       % VIDEO ID =     1137323    
  02:24:24.17 &    -04:32:30.0 &    19.99 &    19.81 &    19.66 &    19.28 &    19.50  &    19.71  &    19.96 &     19.89 &     19.62 &     19.75  &17.35&         -           \\       % VIDEO ID =     1066821    
  02:24:30.91 &    -04:29:57.6 &    22.30  &    22.07 &    21.58 &    21.54 &    21.51 &    21.51  &    21.49 &     21.39 &     20.88 &     20.96   &18.19&        -            \\       % VIDEO ID =     1073325    
  02:24:32.47 &    -04:30:37.2 &    21.21 &    21.24 &    20.85 &    20.80  &    21.11 &    20.77  &    20.86 &     20.94 &     20.72 &     20.47   &5.98&        -           \\       % VIDEO ID =     1071612    
  02:24:32.93 &    -04:47:42.2 &    20.68 &    20.43 &    20.38 &    20.28 &    20.11 &    20.50   &    20.61 &     20.69 &     20.52 &     20.01   &14.95&       -            \\       % VIDEO ID =     1030351    
  02:24:34.34 &    -04:32:00.4 &    20.56 &    19.87 &    19.85 &    19.82 &    19.65 &    19.78  &    19.56 &     19.53 &     19.48 &     19.53  &78.54&       -            \\       % VIDEO ID =     1068133    
  02:24:34.90 &    -04:35:49.9 &    23.28 &    22.40  &    22.22 &    22.11 &    22.23 &    22.49  &    22.30  &     22.25 &     22.20  &     22.13  &-13.70&         -        \\       % VIDEO ID =     1058554    
  02:24:36.91 &    -04:47:09.3 &    22.32 &    21.77 &    21.70  &    21.63 &    21.59 &    21.99  &    21.66 &     21.73 &     21.70  &     21.34  &34.00&         -           \\       % VIDEO ID =     1031542    
  02:24:55.05 &    -04:50:44.0 &    23.81 &    22.72 &    23.04 &    22.94 &    22.67 &    22.83  &    22.72 &     22.63 &     22.29 &     22.23  &-24.33&         -           \\       % VIDEO ID =     1022627    
  02:24:59.17 &    -04:41:52.7 &    21.80  &    21.58 &    21.66 &    21.33 &    21.24 &    21.87  &    22.02 &     21.84 &     21.64 &     21.19  &39.68&        -           \\       % VIDEO ID =     1044760    
  02:25:01.69 &    -04:07:54.4 &    20.09 &    19.87 &    19.61 &    19.54 &    19.64 &    19.82  &    19.56 &     19.71 &     19.58 &     19.43  &1.77&     -            \\       % VIDEO ID =     1126470    
  02:25:02.48 &    -04:05:58.1 &    21.84 &    21.39 &    21.23 &    21.20  &    21.05 &    21.24  &    21.32 &     20.71 &     20.77 &     20.38  &6.88&        -            \\       % VIDEO ID =     1130944    
  02:25:03.60 &    -04:58:28.1 &    21.99 &    21.71 &    21.53 &    21.19 &    21.16 &    21.26  &    21.17 &     21.01 &     20.86 &     20.85  &15.56&         -           \\       % VIDEO ID =     1003005    
  02:25:08.57 &    -04:25:12.8 &    22.13 &    22.01 &    22.04 &    21.71 &    21.62 &    21.80   &    21.92 &     21.95 &     21.60  &     21.19  &-2.54&        -            \\       % VIDEO ID =     1085131    
  02:25:09.55 &    -04:08:38.4 &    20.78 &    20.47 &    20.29 &    20.00  &    20.03 &    20.08  &    20.19 &     20.03 &     19.76 &     19.82  &45.99&       -            \\       % VIDEO ID =     1124634    
  02:25:12.09 &    -04:19:06.9 &    22.55 &    22.46 &    22.18 &    21.98 &    21.86 &    21.98  &    21.90  &     21.87 &     21.76 &     21.40   &26.21&        -            \\       % VIDEO ID =     1099360    
  02:25:14.39 &    -04:47:00.1 &    18.72 &    18.66 &    18.58 &    18.30  &    18.27 &    18.49  &    18.48 &     18.62 &     18.72 &     18.62  &94.18&       -            \\       % VIDEO ID =     1032240    
  02:25:15.35 &    -04:40:08.9 &    20.28 &    19.99 &    19.98 &    19.64 &    19.73 &    19.90   &    19.96 &     20.01 &     20.17 &     19.97  &-4.54&       -            \\       % VIDEO ID =     1048809    
  02:25:16.60 &    -04:05:12.0 &    21.11 &    20.70  &    20.45 &    20.12 &    20.12 &    20.30   &    20.44 &     20.28 &     19.98 &     20.02  &47.26&       -            \\       % VIDEO ID =     1132919    
  02:25:23.28 &    -04:00:38.3 &    21.73 &    21.47 &    21.25 &    21.21 &    21.01 &    21.05  &    21.04 &     21.13 &     21.05 &     20.69  &19.93&         -           \\       % VIDEO ID =     1142883    
  02:25:25.68 &    -04:35:09.6 &    21.24 &    21.02 &    20.92 &    20.72 &    20.62 &    20.90   &    20.85 &     20.86 &     20.67 &     20.25  &360.74&        2.1            \\       % VIDEO ID =     1060199    
  02:25:27.52 &    -03:59:54.4 &    21.87 &    21.76 &    21.66 &    21.22 &    21.26 &    22.61  &    22.48 &     22.27 &     21.79 &     22.14   &27.18&        -            \\       % VIDEO ID =     1143012    
  02:25:31.94 &    -04:06:59.8 &    22.66 &    22.29 &    22.35 &    22.19 &    22.10  &    22.22  &    22.19 &     22.02 &     21.89 &     21.56 &12.48&       -           \\       % VIDEO ID =     1128456    
  02:25:37.16 &    -04:21:32.9 &    20.02 &    19.60  &    19.46 &    19.50  &    19.41 &    19.35  &    19.50  &     19.03 &     19.38 &     19.24  &-21.31&       -           \\       % VIDEO ID =     1094180    
  02:25:38.11 &    -04:28:11.0 &    23.26 &    22.22 &    22.14 &    22.14 &    21.93 &    22.16  &    21.94 &     21.87 &     21.68 &     21.62  &8.21&     -           \\       % VIDEO ID =     1077660    
  02:25:40.57 &    -04:38:25.2 &    20.68 &    20.07 &    20.01 &    19.89 &    19.81 &    20.08  &    19.87 &     19.85 &     19.81 &     19.54        &40.77&         -            \\       % VIDEO ID =     1052734    
  02:25:45.53 &    -04:34:45.6 &    22.06 &    21.55 &    21.57 &    21.20  &    21.29 &    21.36  &    21.44 &     21.24 &     21.02 &     20.94  &8.39&       1.7*            \\       % VIDEO ID =     1061136    
  02:25:50.38 &    -04:33:24.6 &    23.32 &    22.58 &    22.33 &    22.34 &    22.29 &    22.56  &    22.19 &     22.33 &     22.15 &     21.92   &-19.73&         2.7           \\       % VIDEO ID =     1064151    
  02:25:50.97 &    -04:02:47.2 &    22.06 &    21.87 &    21.61 &    21.54 &    21.51 &    21.91  &    21.69 &     21.45 &     21.06 &     20.89         &4.95&       -        \\       % VIDEO ID =     1138410    
  02:25:52.16 &    -04:05:16.1 &    21.76 &    21.83 &    21.56 &    21.40  &    21.42 &    21.43  &    21.40  &     21.19 &     20.80  &     20.72   &45.88&        1.4           \\       % VIDEO ID =     1132702    
  02:25:55.43 &    -04:39:18.3 &    21.35 &    21.12 &    20.82 &    20.73 &    20.62 &    20.92  &    20.79 &     20.56 &     20.38 &     19.95  &47.31&       -            \\       % VIDEO ID =     1050784    
  02:25:57.62 &    -04:50:05.3 &    20.10  &    19.74 &    19.54 &    19.49 &    19.45 &    19.77  &    19.82 &     19.91 &     19.83 &     19.57  &83.74&         -            \\       % VIDEO ID =     1024444    
  02:25:58.87 &    -04:18:01.0 &    22.53 &    22.39 &    22.29 &    22.10  &    21.82 &    22.01  &    21.84 &     21.74 &     21.53 &     21.14  &38.93&         -            \\       % VIDEO ID =     1101783    
  02:26:09.62 &    -04:24:38.0 &    21.83 &    20.91 &    20.80  &    20.83 &    20.62 &    20.63  &    20.43 &     20.45 &     20.50  &     20.45  &17.84&         2.7            \\       % VIDEO ID =     1086539    
  02:26:12.64 &    -04:34:01.4 &    20.82 &    20.84 &    20.81 &    20.62 &    20.39 &    20.67  &    20.67 &     20.62 &     20.43 &     19.86   &9.04&       2.3           \\       % VIDEO ID =     1062840    
  02:26:15.90 &    -04:10:11.3 &    22.17 &    21.87 &    21.56 &    21.51 &    21.59 &    21.47  &    21.55 &     21.55 &     21.35 &     21.22  &14.79&          -            \\       % VIDEO ID =     1120634     
   02:26:17.85 &    -04:31:09.1 &    19.90  &    19.49 &    19.35 &    19.18 &    19.09 &    19.31  &    19.42 &     19.28 &     19.24 &     18.93      &74.06&         -           \\       % VIDEO ID =     1070390   
  
\hline
\label{74quasars}
\end{tabular}

\end{table}
\end{landscape}
\setcounter{table}{0}
\begin{landscape}
\begin{table}

\caption{Positions, AB magnitudes, radio flux densities, and redshifts for the final 74 Gold quasars. $z_{spec}$ is the SALT or VVDS redshift, where available. \newline * This VVDS object is thought to have been assigned an incorrect redshift. The corrected value is given here.}
\begin{tabular}{@{}ccccccccccccrr@{}}
 \hline
R.A. (hms)& Dec. (dms) & $u$ & $g$ & $r$ & $i$ & $z$ & $Z$ & $Y$ & $J$ & $H$ & $K_{{\rm S}}$ & $S_{{\rm 1.4\,GHz}}/\umu$Jy & $z_{spec}$ \\
 \hline

  02:26:18.59 &    -04:11:01.1 &    21.67 &    21.53 &    21.49 &    21.16 &    21.10  &    21.88  &    21.89 &     21.89 &     21.72 &     21.42 &7.52&          2.0           \\       % VIDEO ID =     1118527    
  02:26:19.65 &    -04:46:09.2 &    23.37 &    22.31 &    22.38 &    22.25 &    22.02 &    22.21  &    22.14 &     21.88 &     21.60  &     21.12   &18.84&         -            \\       % VIDEO ID =     1034175    
  02:26:22.15 &    -04:22:21.8 &    19.87 &    19.69 &    19.59 &    19.01 &    18.98 &    19.21  &    19.29 &     19.36 &     19.32 &     18.99   &16.34&        2.0           \\       % VIDEO ID =     1092183    
  02:26:24.64 &    -04:20:02.4 &    21.84 &    21.65 &    21.47 &    21.28 &    21.10  &    21.45  &    21.48 &     21.47 &     21.37 &     20.80   &56.27&        2.2       \\       % VIDEO ID =     1097378    
  02:26:29.26 &    -04:30:57.2 &    20.12 &    19.84 &    19.90  &    19.67 &    19.52 &    19.82  &    19.78 &     19.97 &     20.02 &     19.68  &13.27&         -          \\       % VIDEO ID =     1070872    
  02:26:32.38 &    -04:38:09.0 &    22.88 &    22.48 &    22.45 &    22.04 &    22.05 &    22.35  &    22.22 &     22.05 &     21.80  &     21.36  &-29.17&         -          \\       % VIDEO ID =     1053251    
  02:26:33.31 &    -04:29:47.8 &    22.73 &    22.15 &    22.00  &    21.86 &    21.60  &    21.76  &    21.77 &     21.75 &     21.54 &     21.11 &22.73&         2.1           \\       % VIDEO ID =     1073715    
  02:26:35.48 &    -04:51:04.8 &    20.58 &    20.55 &    20.45 &    20.25 &    20.12 &    20.49  &    20.30  &     20.28 &     20.15 &     19.68  &13.19&       -           \\       % VIDEO ID =     1022014    
  02:26:36.07 &    -04:34:29.1 &    20.82 &    20.00  &    19.77 &    19.59 &    19.63 &    19.60   &    19.48 &     19.50  &     19.60  &     19.67  &33.11&         -            \\       % VIDEO ID =     1061820    
  02:26:39.83 &    -04:20:04.4 &    21.78 &    21.95 &    21.67 &    21.34 &    21.42 &    21.54  &    21.78 &     21.57 &     21.20  &     21.43   &22.22&         1.6            \\       % VIDEO ID =     1097196    
  02:26:41.46 &    -03:59:49.8 &    22.10  &    21.82 &    21.83 &    21.59 &    21.41 &    21.61  &    21.68 &     21.86 &     21.88 &     21.55  &34.03&         -           \\       % VIDEO ID =     1144190    
  02:26:48.19 &    -04:03:14.5 &    23.01 &    22.60  &    22.53 &    22.31 &    22.09 &    22.38  &    22.39 &     22.28 &     21.96 &     21.56        &6.44&        -           \\       % VIDEO ID =     1137348    
  02:26:52.14 &    -04:05:57.1 &    21.22 &    20.95 &    20.66 &    20.51 &    20.46 &    20.35  &    20.55 &     20.31 &     19.93 &     19.84  &31.34&         1.4           \\       % VIDEO ID =     1131001    
  02:26:55.57 &    -04:42:17.9 &    22.60  &    22.26 &    22.16 &    22.04 &    21.73 &    22.10   &    22.06 &     21.98 &     21.76 &     21.22 &6.67&         -         \\       % VIDEO ID =     1043747    
  02:26:57.25 &    -03:59:44.5 &    20.08 &    20.08 &    20.03 &    19.66 &    19.59 &    19.74  &    19.77 &     19.88 &     19.84 &     19.52   &7.89&         -          \\       % VIDEO ID =     1142717    
  02:27:07.54 &    -04:32:03.2 &    21.27 &    20.98 &    20.77 &    20.50  &    20.51 &    20.83  &    20.75 &     20.51 &     20.16 &     20.07 &28.38&         1.5*          \\       % VIDEO ID =     1067992    
  02:27:09.03 &    -04:55:10.1 &    22.19 &    21.51 &    21.25 &    21.08 &    21.09 &    21.08  &    20.95 &     20.97 &     20.96 &     21.14  &19.39&         2.7           \\       % VIDEO ID =     1011226    
  02:27:11.36 &    -04:58:15.1 &    23.03 &    22.75 &    22.51 &    22.25 &    22.14 &    22.89  &    22.76 &     22.26 &     21.80  &     21.61        &-0.82&         -            \\       % VIDEO ID =     1003531    
  02:27:22.19 &    -04:52:21.8 &    21.84 &    21.79 &    21.81 &    21.59 &    21.40  &    21.62  &    21.61 &     21.75 &     21.60  &     21.18  &18.32&        -           \\       % VIDEO ID =     1018510    
  02:27:23.71 &    -04:58:08.3 &    22.79 &    21.95 &    21.86 &    21.70  &    21.38 &    21.79  &    21.58 &     21.49 &     21.24 &     21.19  &13.44&         -          \\       % VIDEO ID =     1003803  
  02:27:23.80 &    -04:02:14.8 &    20.39 &    19.91 &    19.89 &    19.68 &    19.78 &    20.26  &    20.37 &     20.35 &     20.33 &     20.08  &96.93&      -           \\       % VIDEO ID =     1139803    
  02:27:25.72 &    -04:18:37.5 &    23.56 &    22.64 &    22.48 &    22.53 &    22.37 &    22.85  &    22.39 &     22.27 &     22.36 &     22.28   &0.14&        -           \\       % VIDEO ID =     1100418    
  02:27:33.99 &    -04:25:23.5 &    20.64 &    20.50  &    20.40  &    20.13 &    20.21 &    20.32  &    20.42 &     20.22 &     19.91 &     19.90   &14.33&        1.6         \\       % VIDEO ID =     1084813    
  02:27:36.93 &    -04:26:31.5 &    21.87 &    21.53 &    21.49 &    21.16 &    21.29 &    21.35  &    21.31 &     21.21 &     21.13 &     20.94  &23.80&         1.8          \\       % VIDEO ID =     1081826    
  02:27:38.99 &    -04:09:40.8 &    22.19 &    22.00  &    21.71 &    21.64 &    21.53 &    21.61  &    21.62 &     21.43 &     20.97 &     20.85   &-21.24&         1.4           \\       % VIDEO ID =     1122090    
  02:27:40.55 &    -04:02:51.1 &    19.00  &    18.41 &    18.19 &    18.10  &    18.05 &    18.33  &    18.15 &     18.22 &     18.20  &     18.35   &2381.60&         2.6           \\       % VIDEO ID =     1138526    
  02:27:46.46 &    -04:01:00.7 &    22.14 &    22.00  &    21.94 &    21.56 &    21.53 &    21.65  &    21.64 &     21.62 &     21.53 &     21.36  &20.99&         -          \\       % VIDEO ID =     1142963    
  02:27:47.16 &    -04:45:20.1 &    22.13 &    22.02 &    21.83 &    21.34 &    21.43 &    21.72  &    21.49 &     21.36 &     21.27 &     21.06  &72.35&          -          \\       % VIDEO ID =     1036097    
  02:27:47.34 &    -04:27:53.7 &    22.32 &    22.00  &    21.68 &    21.53 &    21.46 &    21.74  &    21.40  &     21.45 &     21.07 &     20.64  &13.33&         1.1           \\       % VIDEO ID =     1078470    
  02:27:49.46 &    -04:59:25.6 &    22.39 &    22.03 &    21.75 &    21.67 &    21.54 &    21.56  &    21.70  &     21.71 &     21.58 &     21.15  &-13.53&        -          \\       % VIDEO ID =     1000617  

\hline
\end{tabular}

\end{table}
\end{landscape}

\label{lastpage}

\end{document}